\documentclass[journal=jpcafh]{achemso}
\setkeys{acs}{articletitle = true}
\SectionNumbersOn
\usepackage[nolist]{acronym}

\usepackage[version=3]{mhchem} 
\usepackage{amsmath,amssymb,amsfonts}
\usepackage{xcolor} 
\usepackage{epsfig}
\usepackage{subfig}
\usepackage{graphicx}
\usepackage{epsfig}
\usepackage{longtable}
\usepackage{bm}
\usepackage{array}
\usepackage{lscape}
\usepackage{siunitx}

\usepackage{subfig}
\usepackage{subfloat}

\usepackage[colorlinks]{hyperref}
\hypersetup{ 
    colorlinks=true,
    linkcolor=red!80!black,
    urlcolor=magenta!80!black,
    citecolor=green!70!black,
}
\usepackage[nameinlink, capitalize]{cleveref}

\usepackage[utf8]{inputenc}
\usepackage[T1]{fontenc}
\usepackage{braket}

\newcommand*{\ee}[1]{\mathrm{e}^{#1}}

\author{H{\aa}kon Emil Kristiansen}
\email{h.e.kristiansen@kjemi.uio.no}
\affiliation[Hylleraas]{Hylleraas Centre for Quantum Molecular Sciences, Department of Chemistry, University of Oslo, P.O. Box 1033 Blindern, N-0315 Oslo, Norway}
\author{H{\aa}kon Kvernmoen}
\affiliation[University of Oslo]{Department of Physics, University of Oslo, P.O. Box 1033 Blindern, N-0315 Oslo, Norway}
\author{Simen Kvaal}
\author{Thomas Bondo Pedersen}
\email{t.b.pedersen@kjemi.uio.no}
\affiliation[Hylleraas]{Hylleraas Centre for Quantum Molecular Sciences, Department of Chemistry, University of Oslo, P.O. Box 1033 Blindern, N-0315 Oslo, Norway}

\title[CC Weights]{Configuration weights in coupled-cluster theory}

\begin{document}


\begin{abstract}    
We introduce a simple definition of the weight of any given Slater determinant in the coupled-cluster state, namely as the expectation value of the projection operator onto that determinant.
The definition can be applied to any coupled-cluster formulation, including conventional coupled-cluster theory, perturbative coupled-cluster models, nonorthogonal orbital-optimized coupled-cluster theory, and extended coupled-cluster theory, allowing for wave-function analyses on par with configuration-interaction-based wave functions.
Numerical experiments show that for single-reference systems the coupled-cluster weights are in excellent
agreement with those obtained from the full configuration-interaction wave function.
Moreover, the well-known insensitivity of the total energy obtained from truncated coupled-cluster models to the choice of orbital basis is clearly exposed by weights computed in the $\hat{T}_1$-transformed determinant basis.
We demonstrate that the inseparability of the conventional linear parameterization of the bra (left state) for systems composed of noninteracting subsystems may lead to ill-behaved (negative or greater than unity) weights, an issue that can only be fully remedied by switching to extended coupled-cluster theory.
The latter is corroborated by results obtained with quadratic coupled-cluster theory, which is shown numerically to yield a significant improvement.
\end{abstract}

\section{Introduction}

The coupled-cluster (CC) method is arguably the most successful and widely used correlated wave-function model in molecular electronic-structure
theory, for excited states as well as for ground states~\cite{Gauss1998,MEST,Crawford2000,Bartlett2007,Shavitt2009,Bartlett2010,bartlett_perspective_2024}
and, in recent years, also for many-electron dynamics~\cite{ofstad_time-dependent_2023} induced by external forces such as ultrashort laser pulses.
The key to its success is the non-unitary exponential parameterization which---in combination with a nonvariational wave-function optimization---yields
an inherently size-extensive and size-consistent hierarchy of wave-function approximations that converges to the formally exact
full configuration-interaction (FCI) theory.

The CC wave function is, however, substantially harder to interpret in elementary quantum-mechanical terms than the FCI wave function.
The latter can be written as a superposition
\begin{equation}
    \label{eq:FCI-WF}
    \ket{\Psi} = \sum_\mu \ket{\Phi_\mu} C_\mu,
\end{equation}
where the summation is over all $N$-electron Slater determinants $\ket{\Phi_\mu}$ that can be constructed with a given set of
orthonormal spin orbitals.
The coefficients $C_\mu$ of the ground-state wave function are computed from the variation principle and
form the eigenvector corresponding to the lowest eigenvalue of the Hamiltonian matrix in the determinant basis.
Evidently, each coefficient $C_\mu = \braket{\Phi_\mu \vert \Psi}$ is the quantum-mechanical
probability amplitude for the system being in the $N$-electron quantum state represented by the Slater determinant $\ket{\Phi_\mu}$.
The normalization condition,
\begin{equation}
    \label{eq:FCI-norm}
    1 = \braket{\Psi \vert \Psi} = \sum_\mu \braket{\Psi \vert \Phi_\mu}\!\braket{\Phi_\mu \vert \Psi} = \sum_\mu \vert C_\mu \vert^2,
\end{equation}
allows one to judge the relative importance of each determinant $\ket{\Phi_\mu}$ in the expansion \eqref{eq:FCI-WF} by its weight
(probability) $\vert C_\mu \vert^2$.
This gives rise to the commonly used terminology of single-reference (a single dominant determinant or configuration) and
multi-reference (multiple significant configurations) wave functions, typically associated with dynamical and nondynamical
electron correlation, respectively.
In time-dependent FCI (TDFCI) theory, the coefficients become explicitly time-dependent and can be related to the population of stationary
states and interference phenomena during the correlated many-electron dynamics.

It must be kept in mind that the weights are not invariant under rotations of the spin-orbital basis and,
therefore, the FCI wave function may appear to be single-reference in one basis but multi-reference in another one spanning the same Hilbert space.
For example, it is well known that the shortest possible expansion is obtained in the FCI natural-orbital basis.~\cite{lowdin_quantum_1955}
Recently, since the FCI natural-orbital basis is unknown in practice,
orbital localization and other unitary transformations have been proposed to compress the 
wave-function expansion in the context of active-space configuration-interaction
theories.~\cite{li_manni_compression_2020,pandharkar_localized_2022,olivares-amaya_ab-initio_2015,baiardi_density_2020}
Although configuration weights depend on the chosen orbital basis and, therefore, generally cannot be used as a strict diagnostic of
single- or multi-reference character of an electronic state, they are practically the only tools available to us for characterizing 
wave functions in terms of electronic configurations. It is, therefore, of interest to define CC configuration weights in a manner that
converges to the FCI limit while being applicable also to those CC approximations for which a wave function is not strictly defined.

The CC wave function is given by
\begin{equation}
    \label{eq:CC-WF}
    \ket{\Psi} = \ee{\hat{T}}\ket{\Phi_0},
\end{equation}
where the cluster operator,
\begin{equation}
    \label{eq:cluster-op}
    \hat{T} = \sum_\mu \tau_\mu \hat{X}_\mu,
\end{equation}
is defined in terms of amplitudes $\tau_\mu$ and excitation operators $\hat{X}_\mu$.
The excitation operators are defined with respect to a chosen reference determinant
$\ket{\Phi_0}$ such that $\ket{\Phi_\mu} = \hat{X}_\mu \ket{\Phi_0}$ and $\braket{\Phi_\mu \vert \Phi_\nu} = \delta_{\mu\nu}$.
The cluster amplitudes $\tau_\mu$ are determined nonvariationally by projection of the Schr{\"o}dinger equation onto the determinant basis
generated by the excitation operators included in the cluster operator.
When the cluster operator is truncated, the reference determinant $\ket{\Phi_0}$ should be chosen as the one dominating the (typically unknown)
FCI expansion in the same orbital basis. If the reference determinant is not dominant,
the truncated CC wave function tend to be a poor approximation unless the single excitations, which
effectively act as orbital relaxation parameters, are able to correct a poorly chosen reference.
The CC wave function is not normalized but
as long as all possible excitations are retained in the cluster operator \eqref{eq:cluster-op}, Eqs.~\eqref{eq:FCI-WF} and \eqref{eq:CC-WF} are
equivalent up to a normalization constant, provided that the reference is not orthogonal
to the exact ground-state wave function.
The main advantage of the exponential parametrization of CC theory is that it conserves
crucial properties of the exact wave function---namely, size consistency and size extensivity~\cite{Bartlett2007,bartlett_perspective_2024}---when
the cluster operator is truncated. These
properties are lost when the expansion \eqref{eq:FCI-WF} is truncated, causing dramatic failures that only grow worse as the system size increases.

Unfortunately, there is no simple quantum-mechanical
interpretation of the cluster amplitudes. It is, of course, possible to compute the overlap of the CC wave function and any Slater determinant,
$\braket{\Phi_\mu \vert \Psi}$, but it cannot be interpreted as a probability amplitude unless the missing normalization is taken into account.
This is effectively the same as mapping the CC wave function onto a FCI wave function with the remarkable result that the CC wave function
has components in the entire $N$-electron space regardless of the truncation level of the cluster operator. Even if the cluster operator is
truncated, the calculation of CC probability amplitudes by mapping onto the FCI wave function scales factorially with $N$ and is, therefore,
(almost) never done in practice.

The fact that the CC wave function has components in the entire $N$-electron space is commonly used
to explain why CC ground-state energies converge faster to the FCI limit than the analogous CI expansions. It should be recalled, however, that
the CC energy,
\begin{equation}
    \label{eq:CC-energy}
    E = \braket{\Phi_0 \vert \hat{H} \left(1 + \hat{T}_1 + \hat{T}_2 + \frac{1}{2} \hat{T}_1^2 \right) \vert \Phi_0},
\end{equation}
only has contributions from the reference, single-excited, and double-excited determinants since the Hamiltonian $\hat{H}$ has excitation
rank $2$ (i.e., is a two-electron operator). Here, the cluster operator is recast as a sum over excitation ranks from $1$ to $N$,
\begin{equation}
    \hat{T} = \sum_{i=1}^N \hat{T}_i = \sum_{i=1}^N \sum_{\mu_i} \tau_{\mu_i} \hat{X}_{\mu_i},
\end{equation}
Equation \eqref{eq:CC-energy} is valid for any truncation of the cluster operator (including after singles where $\hat{T}_2 = 0$) and
the triple and higher-order excitations only affect the energy indirectly through the amplitude equations.

Any other observable is computed with the aid of a dual state defined such that the CC expectation-value
functional fullfils the Hellman-Feynman theorem. While often perceived as nothing but a computationally convenient construction,
the dual state plays an important role for the fundamental physical content of the theory.
This is particularly evident in time-dependent CC (TDCC) theory~\cite{ofstad_time-dependent_2023} which is best formulated in the bivariational framework
of \citeauthor{Arponen1983},~\cite{Arponen1983} effectively mapping the quantum-mechanical problem onto classical Hamiltonian
mechanics.~\cite{Arponen1991,Arponen1997,Pedersen1998} In this formulation, it is clear that $\ket{\Psi}$ and its dual together form
a phase space, indicating that the CC description of a quantum state requires $\emph{both}$. This is also evident from
equation-of-motion CC (EOM-CC)~\cite{Geertsen1989,Stanton1993,Comeau1993} theory where \emph{both} left and right eigenstates are needed
to compute ground- and excited-state properties and transition probabilities.
The relation between TDCC theory and Hamiltonian mechanics was exploited in Ref.~\citenum{Pedersen2019} to propose stable symplectic integration of the TDCC equations
of motion and to guide physical interpretation of the TDCC quantum state using both $\ket{\Psi}$ and its dual on an equal footing. Moreover,
the bivariational viewpoint allows for a simple definition of stationary-state populations as expectation values of
suitable projection operators, thus enabling conventional quantum-mechanical interpretations of TDCC quantum dynamics.~\cite{Pedersen2021}
Analogously, in the present work, we use the bivariational formulation of CC theory to propose expectation-value expressions for the weights $\vert C_\mu \vert^2$,
which are equally valid for truncated cluster operators and at the FCI limit.
This allows for a simple interpretation of the CC state on the same footing as configuration-interaction based wave functions.

\section{Theory}

\subsection{Configuration weights in bivariational theory}

\citeauthor{Arponen1983}'s bivariation principle~\cite{Arponen1983}
is based on \emph{independent} appoximations for the wave function, $\ket{\Psi}$, and its hermitian conjugate, denoted $\bra{\tilde{\Psi}}$,
which are canonical variables analogous to the generalized positions and momenta defining the classical phase space,~\cite{Arponen1991,Arponen1997,Pedersen1998}
and satisfy the normalization condition $\braket{\tilde{\Psi} \vert \Psi} = 1$.
By analogy with the classical phase space, \emph{both} the ket and the bra are needed to represent the quantum state of the $N$-electron system.
In other words, the bra $\bra{\tilde{\Psi}}$ is as physical as the ket $\ket{\Psi}$ and \emph{both} must be taken into account in the
quantum-mechanical interpretation. It is \emph{not} sufficient to consider only the ket $\ket{\Psi}$.
While this is perhaps an unusual viewpoint for ground-state theories, the EOM-CC\cite{Geertsen1989,Stanton1993,Comeau1993}
approach to excited states
operates with ``left'' (bra) and ``right'' (ket) eigenstates, \emph{both} of which are required to compute transition probabilities and
ground- and excited-state properties.~\cite{Stanton1993}

Choosing a particular inner product on the CC phase space,
the expectation-value function becomes~\cite{Pedersen2019}
\begin{equation}
    \label{eq:expect}
    \braket{\hat{O}} = \frac{1}{2} \braket{\tilde{\Psi} \vert \hat{O} \vert \Psi}
                     + \frac{1}{2} \braket{\tilde{\Psi} \vert \hat{O}^\dagger \vert \Psi}^*,
\end{equation}
for some operator $\hat{O}$.
Importantly, the bivariation principle guarantees that this expression fullfils both the 
ordinary time-independent~\cite{Hellmann1937,Feynman1939} and the time-dependent~\cite{Hayes1965} Hellmann-Feynman theorem.~\cite{Arponen1983,Pedersen1998,Pedersen2019}

By analogy with Eq.~\eqref{eq:FCI-norm}, we define the weight $W_\mu$ of a determinant $\ket{\Phi_\mu}$ in the bivariational state as the
expectation value of the projection operator $\hat{P}_\mu = \ket{\Phi_\mu}\!\bra{\Phi_\mu}$,
\begin{equation}
    W_\mu = \braket{\hat{P}_\mu} = \braket{\tilde{\Psi} \vert \hat{P}_\mu \vert \Psi} = \tilde{c}_\mu c_\mu,
\end{equation}
where we have assumed real orbitals and cluster amplitudes, and introduced
\begin{equation}
    \label{eq:c_mu}
    \tilde{c}_\mu = \braket{\tilde{\Psi} \vert \Phi_\mu}, \qquad
    c_\mu = \braket{\Phi_\mu \vert \Psi}.
\end{equation}
By the resolution of the identity, $\sum_\mu \hat{P}_\mu = 1$ where the summation is over \emph{all} $N$-electron Slater determinants
in the given spin-orbital basis, we have
\begin{equation}
    \sum_\mu W_\mu = \braket{\tilde{\Psi} \vert \Psi} = 1,
\end{equation}
which suggests that the bivariational weights may be interpreted in the same way as the FCI weights, i.e., as quantum-mechanical probabilities.
Note, in particular, that one may compute the weights in a \emph{different} Slater-determinant basis than that used to compute the
wave function. 
In general, \emph{any} similarity transformation $\hat{P}_\mu \leftarrow \hat{S}\hat{P}_\mu\hat{S}^{-1}$ can be applied, including unitary orbital rotations,
such that
\begin{equation}
    \label{eq:Wsim}
    W_\mu \leftarrow \braket{\tilde{\Psi} \vert \hat{S}\hat{P}_\mu\hat{S}^{-1} \vert \Psi},
\end{equation}
although doing so may result in intractable computational costs.

One notable caveat arising from the bivariational formulation is that, while inherently real and guaranteed to sum to unity, the individual
weights $W_\mu$ are not bounded below by $0$ nor above by $1$ except at the FCI limit.
The unboundedness is a common feature of non-Hermitian theories and is also present in, e.g.,
EOM-CC theory where transition probabilities may be negative or greater than unity and where
closely related sum rules such as the Thomas-Reiche-Kuhn\cite{thomas_uber_1925,reiche_uber_1925,kuhn_uber_1925} and Condon\cite{Condon1937} sum rules for oscillator strengths 
and rotatory strengths, respectively, are not fulfilled except at the FCI limit (and with a complete orbital basis).~\cite{Pedersen1997,Pedersen1998b,Pedersen1999,Pedersen2004}
Moreover, we note that the same unboundedness also arises in CC stationary-state populations---but
no practical issues were observed in the initial quantum-dynamics studies reported by \citeauthor{Pedersen2021}~\cite{Pedersen2021}

Of particular interest for comparisons between different methods is the reference weight $W_0$ and the total weights of singles, doubles, etc., which we define as
\begin{align}
    &W_1 = \sum_{\mu_1} W_{\mu_1} = \braket{\hat{P}_1}, \\
    &W_2 = \sum_{\mu_2} W_{\mu_2} = \braket{\hat{P}_2},
\end{align}
and so on. Here, we have introduced the total projection operators onto singles, $\hat{P}_1 = \sum_{\mu_1} \hat{P}_{\mu_1}$,
and onto doubles, $\hat{P}_2 = \sum_{\mu_2} \hat{P}_{\mu_2}$. Similar definitions apply for triples ($W_3$), quadruples ($W_4$),
and higher-order excitations.

Related to the important requirement of size-consistency and size-extensivity,
the weights should behave in specific ways when the system is composed of noninteracting subsystems.
For simplicity and without loss of generality, we consider an electronic system composed of two infinitely separated (and hence noninteracting)
subsystems $A$ and $B$.
In FCI theory, the wave function is multiplicatively separable, i.e., $\ket{\Psi} = \ket{\Psi^A}\ket{\Psi^B}$, and expectation values
become either multiplicatively or additively separable when the operator in question is multiplicatively or additively separable, respectively,
see, e.g., Refs.~\citenum{Hansen2019,Hansen2020} for very detailed and general discussions of separability (in the context of vibrational
CC theory). In bivariational theory, ideally, the bra $\bra{\tilde{\Psi}}$ should be multiplicatively separable, too.
Now, the determinant projection operators are not generally separable, neither multiplicatively nor additively, since an excitation
may be either localized on subsystem $A$ or on subsystem $B$, or involve spin orbitals on both subsystems. Assuming that the chosen 
reference determinant is multiplicatively separable, we have the relations
\begin{align}
    &\hat{P}_0 = \hat{P}_0^A \hat{P}_0^B, \\
    &\hat{P}_1 = \hat{P}_1^A \hat{P}_0^B + \hat{P}_0^A \hat{P}_1^B + \hat{P}_1^{AB}, \\
    &\hat{P}_2 = \hat{P}_2^A \hat{P}_0^B + \hat{P}_0^A \hat{P}_2^B + \hat{P}_1^A \hat{P}_1^B + \hat{P}_2^{AB},
\end{align}
for the projection operators onto the reference, singles, and doubles, respectively.
Hence, as long as both $\bra{\tilde{\Psi}}$ and $\ket{\Psi}$ are multiplicatively separable (as in FCI theory), the
corresponding weights can be expressed in terms of subsystem weights according to
\begin{align}
    \label{eq:W0_sep}
    &W_0 = W_0^AW_0^B, \\
    \label{eq:W1_sep}
    &W_1 = W_1^AW_0^B + W_0^AW_1^B, \\
    \label{eq:W2_sep}
    &W_2 = W_2^AW_0^B + W_0^AW_2^B + W_1^AW_1^B.
\end{align}
These relations make it abundantly clear that
weights are not size-extensive quantities and, hence, cannot be used as a rigorous diagnostic for single- or multi-reference character.
At the very least, one would have to use the ratio of the two largest weights, although this measure remains orbital-dependent.
On the other hand, if one observes a reference weight close to unity in a given spin-orbital basis, then the system certainly can be 
characterized as single-reference. For the \ce{He} atom, for example, with the aug-cc-pVDZ basis set and canonical HF spin orbitals,
the FCI reference weight is $W_0 = 0.992$, leaving no doubt that the electronic wave function is single-reference.
However, the wave function would still be single-reference for $860$ noninteracting \ce{He} atoms even though the reference weight would
drop to $W_0 = 0.001$.
More general approaches to the characterization and error assessment of the specific case of CC wave functions have been developed recently.
\citeauthor{Bartlett2020}~\cite{Bartlett2020} introduced size-extensive and orbital-invariant multi-determinant
and multi-reference indices for characterizing CC wave functions,
and \citeauthor{faulstich_s-diagnosticposteriori_2023}~\cite{faulstich_s-diagnosticposteriori_2023} proposed a diagnostic
based on mathematical analysis of CC theory. Still, despite its weaknesses, the weight concept plays a fundamental
role in the understanding of electronic structure and, for example, the dominant weights are commonly
used to describe the wave function obtained from a complete active space self-consistent field calculation in a
given orbital basis.

In the following sections we will discuss weights in the context of various flavors of CC theory.

\subsection{Conventional CC theory}

The most widely employed CC formulation in quantum chemistry uses the parameterization of Eq.~\eqref{eq:CC-WF} with the reference
determinant typically chosen to be the ground-state HF determinant, and
\begin{equation}
    \label{eq:tildePsi}
    \bra{\tilde{\Psi}} = \bra{\Phi_0} (1 + \hat{\Lambda}) \ee{-\hat{T}}.
\end{equation}
Here, the de-excitation cluster operator is defined in terms of amplitudes $\lambda_\mu$ as
\begin{equation}
    \label{eq:Lambda}
    \hat{\Lambda} = \sum_\mu \lambda_\mu \hat{X}_\mu^\dagger,
\end{equation}
where the summation is the same as in the cluster operator, Eq.~\eqref{eq:cluster-op}.
Systematic truncation of the cluster operators lead to a hierarchy of increasingly accurate
models. For example, the CC singles (CCS), CC singles and doubles (CCSD), CC singles doubles and triples (CCSDT) models
are obtained by truncating the cluster operators after singles ($\hat{\Lambda} = \hat{\Lambda}_1$, $\hat{T} = \hat{T}_1$),
after doubles ($\hat{\Lambda} = \hat{\Lambda}_1 + \hat{\Lambda}_2$, $\hat{T} = \hat{T}_1 + \hat{T}_2$), and
after triples ($\hat{\Lambda} = \hat{\Lambda}_1 + \hat{\Lambda}_2 +\hat{\Lambda}_3$, $\hat{T} = \hat{T}_1 + \hat{T}_2 + \hat{T}_3$), respectively.

The bivariation principle requires that the CC Lagrangian (i.e., energy functional) 
\begin{equation}
    \mathcal{L} = \braket{\tilde{\Psi} \vert \hat{H} \vert \Psi},
\end{equation}
be stationary with respect to variations in the amplitudes $\lambda$ and $\tau$.
This leads to the equations
\begin{align}
    \label{eq:tau}
    &\braket{\Phi_\mu \vert \ee{-\hat{T}}\hat{H} \ee{\hat{T}} \vert \Phi_0} = 0, \\
    \label{eq:lambda}
    &\braket{\tilde{\Psi} \vert [\hat{H}, \hat{X}_\mu] \vert \Psi} = 0,
\end{align}
which determine the $\tau$ and $\lambda$ amplitudes.
Note that Eqs.~\eqref{eq:tau} and \eqref{eq:lambda} are uncoupled such that
the $\lambda$ amplitudes can be regarded as functions of the cluster amplitudes $\tau$ and of the Hamiltonian $\hat{H}$.
This is a direct consequence of the linear parametrization of $\hat{\Lambda}$ in Eq.~\eqref{eq:Lambda}.
The $\lambda$ amplitudes can be viewed as Lagrange multipliers that ensure stationarity
of the CC energy $E = \braket{\Phi_0 \vert \exp(-\hat{T}) \hat{H} \vert \Psi}$ under the constraints
defined by Eq.~\eqref{eq:tau}.~\cite{MEST,helgaker_simple_1982,Helgaker1988,Helgaker1992}
The Lagrangian point of view has been demonstrated to yield significant computational advantages
through the so-called $2n+1$ and $2n+2$ rules, which show that the $\tau$ amplitudes to order $n$
in perturbation theory determine the energy through order $2n+1$ while the $\lambda$ amplitudes to
order $n$ determine the energy through order $2n+2$.~\cite{MEST,Helgaker1988,Helgaker1992}
Recently, the Lagrangian technique has been generalized to other properties than the energy,
leading to increased accuracy at significantly reduced computational cost.~\cite{jorgensen_variational_2024}

Alternatively, but equivalently, the linear parameterization of $\hat{\Lambda}$ can be viewed as a 
computationally convenient linear re-parameterization of a de-excitation operator involving the resolvent
of the similarity transformed Hamiltonian, $\bar{H} = \exp(-\hat{T}) \hat{H} \exp(\hat{T})$,
arising from the derivative of Eq.~\eqref{eq:tau} with respect to a perturbation. 
For more details on this formulation, see Ref.~\citenum{bartlett_perspective_2024}
and references therein.

The linear parameterization of $\hat{\Lambda}$ yields a bra, $\bra{\tilde{\Psi}}$, with obvious similarity to configuration-interaction wave functions.
This is unproblematic at the FCI limit (when all excitations are included) but any truncation breaks multiplicative separability
of $\bra{\tilde{\Psi}}$. While expectation values of additively separable operators remain additively separable,
those of multiplicatively separable operators are not multiplicatively separable.~\cite{Hansen2019,Hansen2020}
Since the determinant projection operators are not additively separable (only the reference projector is multiplicatively separable),
the linear parameterization of $\hat{\Lambda}$ implies that truncated CC weights do not obey Eqs.~\eqref{eq:W0_sep}--\eqref{eq:W2_sep}.

Using the definitions \eqref{eq:c_mu}, we may recast $\ket{\Psi}$ and $\bra{\tilde{\Psi}}$ as the configuration-interaction expansions
\begin{equation}
    \bra{\tilde{\Psi}} = \sum_\mu  \tilde{c}_\mu \bra{\Phi_\mu}, \qquad
    \ket{\Psi} = \sum_\mu \ket{\Phi_\mu} c_\mu.
\end{equation}
While the summation in the ket expansion always runs over \emph{all} $N$-electron Slater determinants,
the summation in the bra expansion ends at the truncation level of $\hat{\Lambda}$. Thus,
as a direct consequence of the linear de-excitation operator, CC weights
are only nonzero up to the truncation level of the cluster operators.
For example, for the CCSD model, we have $W_n = 0$ for $n>2$, and $W_0 + W_1 + W_2 = 1$.

The projection operators have excitation rank $0$ in a given spin-orbital basis, prohibiting couplings between
the components of $\bra{\tilde{\Psi}}$ and higher-order components of $\ket{\Psi}$. These do play a role in bivariational CC
theory, however. Expectation values of Hermitian operators with nonzero excitation rank can be written as
\begin{equation}
    \braket{\hat{O}} = \text{Re} \sum_{\mu\nu} \tilde{c}_\mu \braket{\Phi_\mu \vert \hat{O} \vert \Phi_\nu} c_\nu.
\end{equation}
Within CCSD theory, for example, if $\hat{O}$ is a one-electron operator
the doubles components of $\bra{\tilde{\Psi}}$ couple to the triples components of $\ket{\Psi}$. Similarly, for two-electron operators the
quadruples components of $\ket{\Psi}$ contribute.

For the CCS model, doubles and higher-order weights vanish and only the reference and singles weights may be nonzero:
\begin{align}
    \label{eq:W0_CCS}
    &W_0 = 1 - \braket{\Phi_0 \vert \hat{\Lambda}_1\hat{T}_1 \vert \Phi_0}, \\
    \label{eq:Wmu1_CCS}
    &W_{\mu_1} = \braket{\Phi_0 \vert \hat{\Lambda}_1 \vert \Phi_{\mu_1}}\!\braket{\Phi_{\mu_1} \vert \hat{T}_1 \vert \Phi_0}.
\end{align}
Note that if the reference determinant is the HF ground-state wave function, the singles amplitudes vanish.
For the CCSD model, we obtain
\begin{align}
    \label{eq:W0_CCSD}
    &W_0 = 1 - \braket{\Phi_0 \vert \hat{\Lambda}_1\hat{T}_1 \vert \Phi_0} - \braket{\Phi_0 \vert \hat{\Lambda}_2\left(\hat{T}_2 - \frac{1}{2}\hat{T}_1^2\right)\vert \Phi_0}, \\
    \label{eq:Wmu1_CCSD}
    &W_{\mu_1} = \braket{\Phi_0 \vert \hat{\Lambda}_1 \vert \Phi_{\mu_1}}\!\braket{\Phi_{\mu_1} \vert \hat{T}_1 \vert \Phi_0}
               - \braket{\Phi_0 \vert \hat{\Lambda}_2\hat{T}_1 \vert \Phi_{\mu_1}}\!\braket{\Phi_{\mu_1} \vert \hat{T}_1 \vert \Phi_0}, \\
    \label{eq:Wmu2_CCSD}
    &W_{\mu_2} = \braket{\Phi_0 \vert \hat{\Lambda}_2 \vert \Phi_{\mu_2}}\!\braket{\Phi_{\mu_2} \vert \hat{T}_2 + \frac{1}{2}\hat{T}_1^2 \vert \Phi_0},
\end{align}
while the CCSDT weights are given by
\begin{align}
    \label{eq:W0_CCSDT}
    &W_0 = 1 - \braket{\Phi_0 \vert \hat{\Lambda}_1\hat{T}_1 \vert \Phi_0} - \braket{\Phi_0 \vert \hat{\Lambda}_2\left(\hat{T}_2 - \frac{1}{2}\hat{T}_1^2\right)\vert \Phi_0}
    \nonumber \\
    &\qquad - \braket{\Phi_0 \vert \hat{\Lambda}_3\left(\hat{T}_3 - \hat{T}_1\hat{T}_2 + \frac{1}{6}\hat{T}_1^3\right)\vert \Phi_0}, \\
    \label{eq:Wmu1_CCSDT}
    &W_{\mu_1} = \braket{\Phi_0 \vert \hat{\Lambda}_1 \vert \Phi_{\mu_1}}\!\braket{\Phi_{\mu_1} \vert \hat{T}_1 \vert \Phi_0}
               - \braket{\Phi_0 \vert \hat{\Lambda}_2\hat{T}_1 \vert \Phi_{\mu_1}}\!\braket{\Phi_{\mu_1} \vert \hat{T}_1 \vert \Phi_0} \nonumber \\
    &\qquad     - \braket{\Phi_0 \vert \hat{\Lambda}_3\hat{T}_2 \vert \Phi_{\mu_1}}\!\braket{\Phi_{\mu_1} \vert \hat{T}_1 \vert \Phi_0}
                + \frac{1}{2} \braket{\Phi_0 \vert \hat{\Lambda}_3\hat{T}_1^2 \vert \Phi_{\mu_1}}\!\braket{\Phi_{\mu_1} \vert \hat{T}_1 \vert \Phi_0}, \\
    \label{eq:Wmu2_CCSDT}
    &W_{\mu_2} = \braket{\Phi_0 \vert \hat{\Lambda}_2 \vert \Phi_{\mu_2}}\!\braket{\Phi_{\mu_2} \vert \hat{T}_2 + \frac{1}{2}\hat{T}_1^2 \vert \Phi_0}
                - \braket{\Phi_0 \vert \hat{\Lambda}_3\hat{T}_1 \vert \Phi_{\mu_2}}\!\braket{\Phi_{\mu_2} \vert \hat{T}_2 + \frac{1}{2}\hat{T}_1^2 \vert \Phi_0}, \\
    \label{eq:Wmu3_CCSDT}
    &W_{\mu_3} = \braket{\Phi_0 \vert \hat{\Lambda}_3 \vert \Phi_{\mu_3}}\!\braket{\Phi_{\mu_3} \vert \hat{T}_3 + \hat{T}_1\hat{T}_2 + \frac{1}{6}\hat{T}_1^3 \vert \Phi_0}.
\end{align}
Detailed expressions in spin-orbital basis are provided in the appendix.

As is well known, the single-excitation part of the cluster operator, $\hat{T}_1$, acts as an approximate orbital-relaxation operator, making
the CC ground-state energies relatively insensitive to the choice of spin-orbital basis.~\cite{Bartlett2007,bartlett_perspective_2024,Scuseria1987}
At the FCI limit, the CC method becomes fully orbital invariant provided that the chosen reference determinant
is not orthogonal to the FCI wave function.
The effect of single excitations can be elucidated by weights obtained from similarity-transformed projection operators
using Eq.~\eqref{eq:Wsim} with $\hat{S} = \exp(\hat{T}_1)$.
The singles weights vanish identically in this projection basis, whereas the reference weight is expected to increase compared with
the untransformed basis.

\subsection{Alternative formulations}

Since the bivariation principle is based on independent approximations for the bra and ket functions, one might apply alternative
expectation-value functionals based on either $\bra{\tilde{\Psi}}$ or $\ket{\Psi}$ alone, i.e.,
\begin{equation}
    \braket{\hat{O}} = \frac{\braket{\Psi\vert \hat{O} \vert \Psi}}{\braket{\Psi \vert \Psi}} \qquad \mathrm{or}\qquad
    \braket{\hat{O}} = \frac{\braket{\tilde{\Psi}\vert \hat{O} \vert \tilde{\Psi}}}{\braket{\tilde{\Psi}\vert \tilde{\Psi}}}.
\end{equation}
Results computed from either of these expressions will be identical to those computed from Eq.~\eqref{eq:expect} at the
FCI limit. With truncated cluster operators, however, the different expectation-value functionals will produce different
results. For the Hamiltonian, for example, the expectation-value functional should reproduce the CC energy
$E$. However,
\begin{align}
    &\frac{\braket{\Psi\vert \hat{H} \vert \Psi}}{\braket{\Psi \vert \Psi}} =
    E + {\sum_\mu}^\prime \frac{\braket{\Psi\vert \hat{X}_\mu \vert \Psi}}{\braket{\Psi \vert \Psi}} \braket{\Phi_\mu \vert \ee{-\hat{T}} \hat{H} \ee{\hat{T}} \vert \Phi_0}, \\
    &\frac{\braket{\tilde{\Psi}\vert \hat{H} \vert \tilde{\Psi}}}{\braket{\tilde{\Psi}\vert \tilde{\Psi}}} =
    E + {\sum_\mu}^\prime \frac{\braket{\tilde{\Psi}\vert [\hat{H}, \hat{X}_\mu] \vert \Psi}}{\braket{\tilde{\Psi} \vert \tilde{\Psi}}} \braket{\Phi_\mu \vert \ee{-\hat{T}} \vert \tilde{\Psi}}, \\
    &\braket{\tilde{\Psi}\vert \hat{H} \vert \Psi} = E,
\end{align}
where primes indicate summations over excitations not included in the cluster operator (e.g., triples and higher-order excitations for the CCSD model), and where
we have assumed that $\bra{\tilde{\Psi}}$ and $\ket{\Psi}$ are real-valued functions.
At the FCI limit, it follows from Eqs.~\eqref{eq:tau} and \eqref{eq:lambda} that all three expressions yield $E$ but only
the bivariational expectation-value functional reproduces the correct energy with truncated cluster operators.

For configuration weights, the three
expectation-value expressions are identical to leading (i.e., second) order in the amplitudes if one assumes $\hat{\Lambda}_i = \hat{T}_i^\dagger$:
\begin{align}
    &\frac{\braket{\Psi\vert \hat{P}_0 \vert \Psi}}{\braket{\Psi \vert \Psi}} = 1 - \sum_i \braket{\Phi_0 \vert \hat{T}_i^\dagger\hat{T}_i \vert \Phi_0} + \mathcal{O}(\tau^3), \\
    &\frac{\braket{\Psi\vert \hat{P}_{\mu_i} \vert \Psi}}{\braket{\Psi \vert \Psi}} =
    \braket{\Phi_0 \vert \hat{T}_i^\dagger \vert \Phi_{\mu_i}}\!\braket{\Phi_{\mu_i} \vert \hat{T}_i \vert \Phi_0} + \mathcal{O}(\tau^3), \\
    &\frac{\braket{\tilde{\Psi}\vert \hat{P}_0 \vert \tilde{\Psi}}}{\braket{\tilde{\Psi} \vert \tilde{\Psi}}} = 1 - \sum_i \braket{\Phi_0 \vert \hat{\Lambda}_i\hat{\Lambda}_i^\dagger \vert \Phi_0} + \mathcal{O}(z^3), \\
    &\frac{\braket{\tilde{\Psi}\vert \hat{P}_{\mu_i} \vert \tilde{\Psi}}}{\braket{\tilde{\Psi} \vert \tilde{\Psi}}} =
    \braket{\Phi_0 \vert \hat{\Lambda}_i \vert \Phi_{\mu_i}}\!\braket{\Phi_{\mu_i} \vert \hat{\Lambda}_i^\dagger \vert \Phi_0} + \mathcal{O}(z^3), \\
    &\braket{\tilde{\Psi}\vert \hat{P}_0 \vert \Psi} = 1 - \sum_i \braket{\Phi_0 \vert \hat{\Lambda}_i\hat{T}_i \vert \Phi_0} + \mathcal{O}(z^3), \\
    &\braket{\tilde{\Psi}\vert \hat{P}_{\mu_i} \vert \Psi} =
    \braket{\Phi_0 \vert \hat{\Lambda}_i \vert \Phi_{\mu_i}}\!\braket{\Phi_{\mu_i} \vert \hat{T}_i \vert \Phi_0} + \mathcal{O}(z^3),
\end{align}
where $z$ denotes $\lambda$ and $\tau$ amplitudes collectively, and the summations are over the excitation ranks included in the cluster operators.
Thus, to leading order in the amplitudes, configuration weights above the truncation level of the cluster operators vanish with either of the
three expressions.

The $\ket{\Psi}$ and $\bra{\tilde{\Psi}}$ expectation-value functionals yield weights that are bounded below by $0$ and above by $1$ regardless of the
truncation level of the cluster operators. The former can be computed from $\tau$ amplitudes alone, while the latter also requires the $\lambda$ amplitudes.
The weights obtained from $\ket{\Psi}$ are generally nonzero in the entire $N$-body Hilbert space, whereas the $\bra{\tilde{\Psi}}$ weights are nonzero only
for excitations within the truncation level of the cluster operators. Thus, computing weights from $\ket{\Psi}$ alone has FCI complexity regardless of the truncation,
necessitating approximations such as, e.g., truncating the linear re-expansion of $\ket{\Psi}$ at some chosen excitation level. This is unfortunate since the
full $\ket{\Psi}$ expectation-value functional is required to ensure the correct separability properties regardless of the cluster-operator truncation.

More importantly, only the bivariational expectation-value functional is in agreement with the Hellmann-Feynman theorem at any truncation level. This makes it
preferable over the other two expressions for the calculation of ground-state properties in CC theory, including configuration weights. As we shall see below, this choice
also allows us to define configuration weights for perturbation theories where a wave function is not explicitly defined.
Finally, as discussed by \citeauthor{Stanton1993},~\cite{Stanton1993} we stress that the bivariational expectation-value functional emerges
naturally from EOM-CC theory and thus allows us to define configuration weights for excited states as well as the ground state within a single common framework.

\subsection{CC perturbation theories}

Some of the most widely used CC methods are based on perturbation theory and, as such, do not involve an explicit wave-function
parameterization. Examples include the popular second-order M{\o}ller-Plesset (MP2)\cite{Moller1934,MEST} theory and the
related second-order approximation to CCSD theory, the CC2 model,~\cite{Christiansen1995} and the fourth-order approximation to full triples treatment,
the CC3 model,~\cite{koch_cc3_1997} which is often considered to be of benchmark quality, especially for response properties and excitation
energies.~\cite{Schreiber2008} Also the ``Gold Standard'' method of quantum chemistry, the CCSD method with perturbative connected triples
correction (CCSD(T)),\cite{Raghavachari1989} belongs to the set of approximations that do not provide an explicit wave-function expression.

Even in the absence of explicit wave-function expressions, one can still use the expectation-value approach.
One simply starts from the bivariational energy functional and
defines the expectation-value functional in agreement with the Hellman-Feynman theorem. Replacing the Hamiltonian operators with
projection operators then leads to configuration weights for such perturbative CC methods.

For the CC2 model, the energy functional is given by~\cite{Christiansen1995}
\begin{equation}
    \mathcal{L} = \braket{\Phi_0 \vert \left(1 + \hat{\Lambda}_1\right)\left(H + [H,\hat{T}_2]\right) \vert \Phi_0}
    + \braket{\Phi_0 \vert \hat{\Lambda}_2 \left( H + [F,\hat{T}_2] \right) \vert \Phi_0},
\end{equation}
where we have introduced the notation
\begin{equation}
    O = \ee{-\hat{T}_1} \hat{O} \ee{\hat{T}_1},
\end{equation}
for $\hat{T}_1$-transformed operators, $\hat{F}$ is the Fock operator, and $\ket{\Phi_0}$ is the canonical HF ground-state determinant.
Replacing $\hat{H}$ and $\hat{F}$ with projection operators, we obtain the same expressions for the
reference, singles, and doubles weights as for the CCSD model above, Eqs.~\eqref{eq:W0_CCSD}--\eqref{eq:Wmu2_CCSD}.
The only difference is that the amplitudes are evaluated from the CC2 equations rather than the CCSD ones.

Similarly, the CC3 energy functional is defined as~\cite{koch_cc3_1997}
\begin{align}
    \mathcal{L} &= \braket{\Phi_0 \vert H + [H,\hat{T}_2] \vert \Phi_0}
    + \braket{\Phi_0 \vert \hat{\Lambda}_1 \left( H + [H,\hat{T}_2] + [H,\hat{T}_3] \right) \vert \Phi_0} \nonumber \\
    &+ \braket{\Phi_0 \vert \hat{\Lambda}_2 \left( H + [H,\hat{T}_2] + \frac{1}{2}[[H,\hat{T}_2],\hat{T}_2] + [H,\hat{T}_3] \right) \vert \Phi_0} \nonumber \\
    &+ \braket{\Phi_0 \vert \hat{\Lambda}_3 \left( [H,\hat{T}_2] + [F,\hat{T}_3] \right) \vert \Phi_0},
\end{align}
from which one easily obtains weights by replacing $\hat{H}$ and $\hat{F}$ with projection operators.
The resulting expressions for the weights differ from the CCSDT ones in Eqs.~\eqref{eq:W0_CCSDT}--\eqref{eq:Wmu3_CCSDT} and are given by
\begin{align}
    \label{eq:W0_CC3}
    &W_0 = 1 - \braket{\Phi_0 \vert \hat{\Lambda}_1\hat{T}_1 \vert \Phi_0} - \braket{\Phi_0 \vert \hat{\Lambda}_2\left(\hat{T}_2 - \frac{1}{2}\hat{T}_1^2\right)\vert \Phi_0}
    - \braket{\Phi_0 \vert \hat{\Lambda}_3\left(\hat{T}_3 - \hat{T}_1\hat{T}_2\right)\vert \Phi_0}, \\
    \label{eq:Wmu1_CC3}
    &W_{\mu_1} = \braket{\Phi_0 \vert \hat{\Lambda}_1 \vert \Phi_{\mu_1}}\!\braket{\Phi_{\mu_1} \vert \hat{T}_1 \vert \Phi_0}
               - \braket{\Phi_0 \vert \hat{\Lambda}_2\hat{T}_1 \vert \Phi_{\mu_1}}\!\braket{\Phi_{\mu_1} \vert \hat{T}_1 \vert \Phi_0} \nonumber \\
    &\qquad     - \braket{\Phi_0 \vert \hat{\Lambda}_3\hat{T}_2 \vert \Phi_{\mu_1}}\!\braket{\Phi_{\mu_1} \vert \hat{T}_1 \vert \Phi_0}, \\
    \label{eq:Wmu2_CC3}
    &W_{\mu_2} = \braket{\Phi_0 \vert \hat{\Lambda}_2 \vert \Phi_{\mu_2}}\!\braket{\Phi_{\mu_2} \vert \hat{T}_2 + \frac{1}{2}\hat{T}_1^2 \vert \Phi_0}
                - \braket{\Phi_0 \vert \hat{\Lambda}_3\hat{T}_1 \vert \Phi_{\mu_2}}\!\braket{\Phi_{\mu_2} \vert \hat{T}_2 \vert \Phi_0}, \\
    \label{eq:Wmu3_CC3}
    &W_{\mu_3} = \braket{\Phi_0 \vert \hat{\Lambda}_3 \vert \Phi_{\mu_3}}\!\braket{\Phi_{\mu_3} \vert \hat{T}_3 + \hat{T}_1\hat{T}_2 \vert \Phi_0}.
\end{align}
Compared with CCSDT, the missing terms in the CC3 weights are those that involve $\hat{\Lambda}_3$ in conjunction with higher-order singles which are
removed in the perturbation expansion defining the CC3 model.

The CC2 and CC3 models are mainly aimed at time- or frequency-dependent properties such as dynamic polarizabilities and hyperpolarizabilities. For ground-state
energies, they usually can be replaced by non-iterative perturbation theories such as MP2 and CCSD(T), respectively. For the MP2 model, 
the energy functional is defined by
\begin{equation}
    \mathcal{L} = \braket{\Phi_0 \vert \hat{H} + [\hat{H},\hat{T}_2] \vert \Phi_0}
    + \braket{\Phi_0 \vert \hat{\Lambda}_2 \left( \hat{H} + [\hat{F},\hat{T}_2] \right) \vert \Phi_0},
\end{equation}
leading to the weights
\begin{align}
    \label{eq:W0_MP2}
    &W_0 = 1 - \braket{\Phi_0 \vert \hat{\Lambda}_2\hat{T}_2 \vert \Phi_0}, \\
    \label{eq:Wmu2_MP2}
    &W_{\mu_2} = \braket{\Phi_0 \vert \hat{\Lambda}_2 \vert \Phi_{\mu_2}}\!\braket{\Phi_{\mu_2} \vert \hat{T}_2 \vert \Phi_0}.
\end{align}
The singles weights vanish ($W_{\mu_1} = 0$) and $\hat{\Lambda}_2 = \hat{T}_2^\dagger$ in MP2 theory.
The CCSD(T) energy functional can be written as~\cite{Hald2003}
\begin{align}
    \mathcal{L} &= \braket{\Phi_0\vert (1 + \hat{\Lambda}_1 + \hat{\Lambda}_2) \ee{-\hat{T}_1-\hat{T}_2} \hat{H} \ee{\hat{T}_1+\hat{T}_2} \vert \Phi_0}
    + \sum_{i=1}^2 \sum_{\mu_i} \tau_{\mu_i} \braket{\Phi_{\mu_i} \vert [\hat{H}, \hat{T}_3] \vert \Phi_0} \nonumber \\
    \label{eq:L_CCSD(T)}
    &+ \braket{\Phi_0 \vert \hat{\Lambda}_3 \left( [\hat{F},\hat{T}_3] + [\hat{H},\hat{T}_2] \right) \vert \Phi_0},
\end{align}
from which we obtain
\begin{align}
    \label{eq:W0_CCSD_T}
    &W_0 = 1 - \braket{\Phi_0 \vert \hat{\Lambda}_1\hat{T}_1 \vert \Phi_0} - \braket{\Phi_0 \vert \hat{\Lambda}_2\left(\hat{T}_2 - \frac{1}{2}\hat{T}_1^2\right)\vert \Phi_0}
     - \braket{\Phi_0 \vert \hat{\Lambda}_3\hat{T}_3 \vert \Phi_0}, \\
    \label{eq:Wmu1_CCSD_T}
    &W_{\mu_1} = \braket{\Phi_0 \vert \hat{\Lambda}_1 \vert \Phi_{\mu_1}}\!\braket{\Phi_{\mu_1} \vert \hat{T}_1 \vert \Phi_0}
               - \braket{\Phi_0 \vert \hat{\Lambda}_2\hat{T}_1 \vert \Phi_{\mu_1}}\!\braket{\Phi_{\mu_1} \vert \hat{T}_1 \vert \Phi_0}, \\
    \label{eq:Wmu2_CCSD_T}
    &W_{\mu_2} = \braket{\Phi_0 \vert \hat{\Lambda}_2 \vert \Phi_{\mu_2}}\!\braket{\Phi_{\mu_2} \vert \hat{T}_2 + \frac{1}{2}\hat{T}_1^2 \vert \Phi_0}, \\
    \label{eq:Wmu3_CCSD_T}
    &W_{\mu_3} = \braket{\Phi_0 \vert \hat{\Lambda}_3 \vert \Phi_{\mu_3}}\!\braket{\Phi_{\mu_3} \vert \hat{T}_3 \vert \Phi_0}.
\end{align}
The expressions for the MP2 weights are identical to those obtained in CCD theory (the CCSD expressions with $\hat{T}_1 = 0$).
The CCSD(T) weight expressions, on the other hand, differ from both CC3 and full CCSDT by the lack of \emph{all} disconnected triples contributions.
Since the CCSD(T) model consists of an energy correction from perturbative connected triples only, the expressions for the singles and doubles weights
are identical to those obtained from CCSD theory. The computed singles and doubles weights are different, however, since the $\lambda_{\mu_1}$ and
$\lambda_{\mu_2}$ amplitudes are affected by the perturbative triples corrections in Eq.~\eqref{eq:L_CCSD(T)}.

\subsection{Nonorthogonal orbital-optimized CC theory}

Orbital relaxation can be included explicitly in the CC formulation by replacing $\exp(\hat{T}_1)$ with an orbital-rotation
operator $\exp(\hat{\kappa})$, where
\begin{equation}
    \hat{\kappa} = \sum_{\mu_1} \left( \kappa^{e}_{\mu_1} \hat{X}_{\mu_1} + \kappa^{d}_{\mu_1}  \hat{X}_{\mu_1}^\dagger \right).
\end{equation}
The bivariational CC \emph{Ansatz} then becomes
\begin{align}
    &\ket{\Psi} = \ee{\hat{\kappa}} \ee{\hat{T}} \ket{\Phi_0}, \\
    &\bra{\tilde{\Psi}} = \bra{\Phi_0} (1 + \hat{\Lambda}) \ee{-\hat{T}} \ee{-\hat{\kappa}},
\end{align}
where singles are excluded from the cluster operators $\hat{T}$ and $\hat{\Lambda}$. By restricting $\hat{\kappa}$ to be anti-Hermitian,
$\hat{\kappa}^\dagger = -\hat{\kappa}$, such that the orbital-rotation operator is unitary, we obtain the orbital-optimized CC (OCC)
model.~\cite{Purvis1982,Scuseria1987,Sherrill1998,Pedersen1999,Sato2018}
The OCC model, however, fails to converge to the FCI limit for systems with more than two electrons.~\cite{Kohn2005}
As demonstrated by \citeauthor{Myhre2018},~\cite{Myhre2018}
this issue can be removed by lifting the anti-Hermiticity restriction on $\hat{\kappa}$, yielding the nonorthogonal orbital-optimized
CC (NOCC) theory~\cite{Pedersen2001,Kristiansen2020} (or its active-space generalization coined orbital-adaptive time-dependent CC (OATDCC)~\cite{Kvaal2012} theory).

The NOCC equations are identical to the conventional CC equations \eqref{eq:tau} and \eqref{eq:lambda} with singles amplitudes removed and with the
Hamiltonian replaced by the similarity-transformated operator $\hat{H} \leftarrow \exp(-\hat{\kappa})\hat{H}\exp(\hat{\kappa})$,
while the orbital-rotation parameters $\kappa^e$ and $\kappa^d$ are determined by generalized Brillouin conditions.~\cite{Pedersen2001}
The four sets of equations are coupled and must be solved simultaneously---i.e., the $\lambda$ amplitudes are no longer given as
functions of the $\tau$ amplitudes and, hence, cannot be viewed as Lagrangian multipliers.

The NOCC configuration weights can easily be computed from Eq.~\eqref{eq:Wsim} with $\hat{S}=\exp(\hat{\kappa})$,
which implies that singles weights are identically zero. Projection onto the untransformed Slater determinants---typically chosen to
be HF determinants---is not generally feasible, as it would require a computational effort comparable to a FCI calculation.
Truncating the cluster operators after double excitations gives the NOCC doubles (NOCCD) model for which weights can
be computed using the CCSD expressions in Eqs.~\eqref{eq:W0_CCSD} and \eqref{eq:Wmu2_CCSD} with $\hat{\Lambda}_1 = \hat{T}_1 = 0$ 
(see the appendix for full detail).
Note that weights beyond doubles vanish in NOCCD theory since the de-excitation cluster operator $\hat{\Lambda}$ remains linear in truncated NOCC theory.
By the same token, $\bra{\tilde{\Psi}}$ is not multiplicatively separable in truncated NOCC theory and, hence,
the NOCC weights do not obey Eqs.~\eqref{eq:W0_sep} and \eqref{eq:W2_sep} for noninteracting subsystems.

\subsection{Quadratic CC theory}

The only generally applicable way to ensure separability of $\bra{\tilde{\Psi}}$ is to replace the linear de-excitation cluster operator
with an exponential operator,
\begin{align}
    &\ket{\Psi} = \ee{\hat{T}}\ket{\Phi_0}, \\
    \label{eq:ECC_bra}
    &\bra{\tilde{\Psi}} = \bra{\Phi_0} \ee{\hat{\Sigma}} \ee{-\hat{T}},
\end{align}
where $\hat{\Sigma} = \sum_\mu \sigma_\mu \hat{X}_\mu^\dagger$, including singles in both $\hat{T}$ and $\hat{\Sigma}$.
This \emph{Ansatz} defines extended CC (ECC) theory, which was proposed and analyzed in detail by Arponen and coworkers.~\cite{Arponen1983,Arponen1987,Arponen1987a}
The ECC equations are significantly more complicated and computationally demanding than the conventional CC equations and, therefore, applications have been
scarce.~\cite{fan_non-iterative_2005,cooper_benchmark_2010,evangelista_alternative_2011,joshi_extended_2012,Joshi2014,laestadius_analysis_2018,Kvaal2020}
Multiplicative separability at any truncation level of $\bra{\tilde{\Psi}}$ as well $\ket{\Psi}$ was explicitly demonstrated by \citeauthor{Hansen2020}~\cite{Hansen2020}
Their work aimed at vibrational ECC theory but applies to electronic systems as well. Hence, the weights in truncated (as well as untruncated) ECC theory
behave correctly for noninteracting subsystems.

Rather than the full ECC method, we will in the present work consider the quadratic CC (QCC)~\cite{VanVoorhis2000,byrd_quadratic_2002}
method obtained by expanding the exponential de-excitation operator in Eq.~\eqref{eq:ECC_bra} to second order, i.e.,
\begin{equation}
    \label{eq:QCC_bra}
    \bra{\tilde{\Psi}} \leftarrow \bra{\Phi_0} (1 + \hat{\Sigma} + \frac{1}{2}\hat{\Sigma}^2) \ee{-\hat{T}}.
\end{equation}
Truncation after doubles yields the QCC singles and doubles (QCCSD) model, which includes up to quadruple de-excitations through the quadratic term in Eq.~\eqref{eq:QCC_bra}.
Hence, up to quadruple-excitation weights are nonzero in QCCSD theory:
\begin{align}
    \label{eq:W0_QCCSD}
    W_0 &= 1 - \braket{\Phi_0 \vert \hat{\Sigma}_1\hat{T}_1\vert \Phi_0}
    - \braket{\Phi_0 \vert \left(\hat{\Sigma}_2 + \frac{1}{2}\hat{\Sigma}_1^2 \right)\left( \hat{T}_2 - \frac{1}{2}\hat{T}_1^2\right)\vert \Phi_0} \nonumber \\
    &- \braket{\Phi_0 \vert \hat{\Sigma}_1\hat{\Sigma}_2 \left(\frac{1}{6}\hat{T}_1^3 - \hat{T}_1\hat{T}_2 \right) \vert \Phi_0}
    + \frac{1}{4} \braket{\Phi_0 \vert \hat{\Sigma}_2^2 \left( \hat{T}_2^2 - \hat{T}_1^2\hat{T}_2 + \frac{1}{12}\hat{T}_1^4 \right) \vert \Phi_0}, \\
    W_{\mu_1} &= \braket{\Phi_0 \vert \hat{\Sigma}_1 \vert \Phi_{\mu_1}}\!\braket{\Phi_{\mu_1} \vert \hat{T}_1\vert \Phi_0}
    - \braket{\Phi_0 \vert \left(\hat{\Sigma}_2 + \frac{1}{2}\hat{\Sigma}_1^2 \right) \hat{T}_1 \vert \Phi_{\mu_1}}\!\braket{\Phi_{\mu_1} \vert \hat{T}_1\vert \Phi_0} \nonumber \\
    &- \braket{\Phi_0 \vert \hat{\Sigma}_1\hat{\Sigma}_2 \left(\hat{T}_2 - \frac{1}{2}\hat{T}_1^2\right) \vert \Phi_{\mu_1}}\!\braket{\Phi_{\mu_1} \vert \hat{T}_1\vert \Phi_0} \nonumber \\
    &- \frac{1}{2}  \braket{\Phi_0 \vert \hat{\Sigma}_2^2 \left( \frac{1}{6}\hat{T}_1^3 - \hat{T}_1\hat{T}_2 \right)\vert \Phi_{\mu_1}}\!\braket{\Phi_{\mu_1} \vert \hat{T}_1\vert \Phi_0}, \\
    W_{\mu_2} &= \braket{\Phi_0 \vert \hat{\Sigma}_2 + \frac{1}{2}\hat{\Sigma}_1^2 \vert \Phi_{\mu_2}}\!\braket{\Phi_{\mu_2} \vert \hat{T}_2 + \frac{1}{2}\hat{T}_1^2\vert \Phi_0}
    - \braket{\Phi_0 \vert \hat{\Sigma}_1\hat{\Sigma}_2\hat{T}_1\vert  \Phi_{\mu_2}}\!\braket{\Phi_{\mu_2} \vert \hat{T}_2 + \frac{1}{2}\hat{T}_1^2\vert \Phi_0} \nonumber \\
    &- \frac{1}{2} \braket{\Phi_0 \vert \hat{\Sigma}_2^2 \left(\hat{T}_2 - \frac{1}{2}\hat{T}_1^2\right) \vert  \Phi_{\mu_2}}\!\braket{\Phi_{\mu_2} \vert \hat{T}_2 + \frac{1}{2}\hat{T}_1^2\vert \Phi_0}, \\
    W_{\mu_3} &= \braket{\Phi_0 \vert \hat{\Sigma}_1\hat{\Sigma}_2 \vert \Phi_{\mu_3}}\!\braket{\Phi_{\mu_3} \vert \hat{T}_1\hat{T}_2 + \frac{1}{6}\hat{T}_1^3 \vert \Phi_0}
    - \frac{1}{2} \braket{\Phi_0 \vert \hat{\Sigma}_2^2\hat{T}_1 \vert \Phi_{\mu_3}}\!\braket{\Phi_{\mu_3} \vert \hat{T}_1\hat{T}_2 + \frac{1}{6}\hat{T}_1^3 \vert \Phi_0}, \\
    W_{\mu_4} &= \frac{1}{4}
    \braket{\Phi_0 \vert \hat{\Sigma}_2^2\vert \Phi_{\mu_4}}\!\braket{\Phi_{\mu_4} \vert \hat{T}_2^2 + \hat{T}_1^2\hat{T}_2 + \frac{1}{12}\hat{T}_1^4 \vert \Phi_0}.
\end{align}
Detailed expressions for the reference, singles, doubles, triples, and quadruples weights in spin-orbital basis
can be found in Ref.~\citenum{kvernmoen_quadratic_2024} along with the working equations for determining the $\tau$ and $\sigma$ amplitudes.

The truncation of the exponential in Eq.~\eqref{eq:QCC_bra} implies that $\bra{\tilde{\Psi}}$ is not multiplicatively separable. Nevertheless, the inclusion of
the quadratic term is expected to reduce the deviation from separability compared with conventional CCSD theory, especially for four-electron systems where quadruple
de-excitations will be important.

\section{Results}

\subsection{Computational details}

Calculations were performed with the PySCF~\cite{PySCF} and HyQD~\cite{HyQD} program packages using closed-shell spin-restricted implementations
of HF and Kohn-Sham (KS) density-functional theory. For the latter, we used the
Tao-Perdew-Staroverov-Scuseria hybrid density functional (TPSS0)~\cite{tao_climbing_2003,perdew_meta-generalized_2004}
with $25\%$ HF exchange, as implemented in the libxc software library.~\cite{libxc}
All electrons were correlated unless stated otherwise. Both the correlation-consistent double- and triple-zeta basis sets cc-pVDZ and cc-pVTZ were
used.~\cite{dunning_gaussian_1989,prascher_gaussian_2011}
In a few cases, we also used the 6-31G basis set.~\cite{hehre_selfconsistent_1972} All basis set definitions are taken from the \textit{Basis Set Exchange}.~\cite{Feller1996,Schuchardt2007,pritchard2019new}

\subsection{Validation of the CC weight concept}

We start by comparing the weights obtained from the conventional CCSD method with 
those obtained from FCI theory, using the restricted HF (RHF) reference determinant in both cases.
Table \ref{tab:atoms} lists the reference, singles, and doubles weights for the atoms \ce{He}, \ce{Be}, \ce{Ne}, and \ce{Ar} obtained
with the CCSD method and their difference with respect to the FCI results, $\Delta W_n = W_n^\text{CCSD} - W_n^\text{FCI}$, along
with the energy difference, $\Delta E = E^\text{CCSD} - E^\text{FCI}$.
\begin{table*}[htb]
    \centering
    \caption{CCSD reference, singles, and doubles weights for selected closed-shell atoms and errors relative to FCI results.
             The \ce{Ne} core of the \ce{Ar} atom is kept frozen in the correlation treatment. Energy differences are given
             in $\text{mE}_\text{h}$.\label{tab:atoms}}
    \begin{tabular}{l l S[table-format=1.1] S[table-format=1.5] S[table-format=1.5] l S[table-format=1.5] S[table-format=1.5] S[table-format=1.5] S[table-format=2.5]}
    \hline
    \hline
        Atom & {Basis set} & {$\Delta E$} & {$W_0^{\text{CCSD}}$} & {$\Delta W_0$} & {$W_1^{\text{CCSD}}$} & {$\Delta W_1$} & {$W_2^{\text{CCSD}}$} & {$\Delta W_2$} \\ 
    \hline
    \ce{He}  & cc-pVTZ & 0.0 & 0.99216 & 0.00000 & 0.00001 & 0.00000 & 0.00784 &  0.00000 \\
   	\ce{Be}  & cc-pVTZ & 0.3 & 0.90817 & 0.00096 & 0.00143 & 0.00000 & 0.09040 & -0.00090 \\
    \ce{Ne}  & cc-pVDZ & 1.2 & 0.97256 & 0.00022 & 0.00004 & 0.00000 & 0.02740 &  0.00026 \\
	\ce{Ar}  & cc-pVDZ & 1.5 & 0.95149 & 0.00047 & 0.00001 & 0.00000 & 0.04850 &  0.00067 \\
   	\hline
    \hline
    \end{tabular}
\end{table*}
As expected, the CCSD and FCI results are identical (to within convergence thresholds) for the \ce{He} atom, which is evidently a single-reference problem
with a reference weight of $99.2\%$, essentially no singles weight, and $0.8\%$ doubles weight. Also the \ce{Ne} and \ce{Ar} atoms are clear-cut single-reference
problems, with reference weight above $95\%$ and less than $5\%$ doubles weight, in excellent agreement with FCI theory where higher-order excited determinants
are negligible.

The agreement with FCI theory is only slightly worse for the \ce{Be} atom, which has about $9\%$ doubles weight and $91\%$ reference weight.
The CCSD method predicts that two doubly-excited configurations contribute significantly to $W_2$ in this case, $\ket{1\text{s}^22\text{p}^2}$ with weight $0.044$
($49.10 \%$ of $W_2$) and $\ket{1\text{s}^22\text{p}3\text{p}}$ with weight $0.035$ ($38.99\%$ of $W_2$),
in good agreement with the FCI weights $0.045$ ($49.11\%$ of $W_2$) and $0.036$ ($39.04\%$ of $W_2$), respectively.
Using the restricted KS (RKS) orbital basis instead of the RHF one leads to a CCSD energy decrease by just $2.7\,\mu\text{E}_\text{h}$ ($7.1\,\text{J/mol}$).
The reference, singles, and doubles weights are virtually unchanged but
the distribution of doubles weight between the $\ket{1\text{s}^22\text{p}^2}$ and $\ket{1\text{s}^22\text{p}3\text{p}}$ configurations is changed to
$77\%$ and $16\%$, respectively.

More validation data can be found in Tables \ref{tab:mp2_cc2_ccsd_fci_H2}-\ref{tab:N2_ccsd_qccsd_fci} for diatomic molecules at different internuclear distances.
\begin{table*}[htb]
\centering
\caption{Reference, singles, and doubles weights for the \ce{H2} molecule obtained with the cc-pVTZ basis set. The equilibrium
    bond distance is $R_e=1.4\,\text{a}_0$. Energy differences are reported in $\text{mE}_\text{h}$ and the FCI energies are
    $-1.17233459\,\text{E}_\text{h}$ at $R_e$, $-1.01096374\,\text{E}_\text{h}$ at $3R_e$, and $-0.99963751\,\text{E}_\text{h}$ at $6R_e$.
    \label{tab:mp2_cc2_ccsd_fci_H2}}
\begin{tabular}{l S[table-format=1.1] S[table-format=1.7] S[table-format=1.7] S[table-format=1.7] S[table-format=1.7] }
\hline 
\hline 
      &{$R/R_e$}& {MP2}  	& {CC2}       & {CCSD}        & {FCI} \\
\hline
$W_0$ & 1  	    &  0.98987      &  0.98975 	  &  0.98209      &  0.98209       \\
      & 3  	    &  0.92553      &  0.90779 	  &  0.71195      &  0.71195       \\
      & 6  	    &  0.30763      &  0.09465 	  &  0.48444      &  0.48444       \\
$W_1$ & 1  	    &  0.00000      &  0.00009 	  &  0.00012 	  &  0.00012 	   \\
      & 3  	    &  0.00000      &  0.00715 	  &  0.01474      &  0.01474       \\
      & 6  	    &  0.00000      &  0.04904 	  &  0.02355      &  0.02355       \\
$W_2$ & 1  	    &  0.01013      &  0.01015 	  &  0.01779 	  &  0.01779 	   \\
      & 3  	    &  0.07447      &  0.08506 	  &  0.27331      &  0.27331       \\
      & 6  	    &  0.69237      &  0.85631 	  &  0.49201      &  0.49201       \\
\hline
$\Delta E$ & 1  & 7.695         &  7.601      &  0.000        &  \\
		   & 3  & 53.562        & 49.204      &  0.000        &  \\
		   & 6  & 28.153        & -1.073      &  0.000        &  \\
\hline  
\hline
\end{tabular}
\end{table*}

\begin{table*}[htb]
\centering
\caption{Reference, singles, doubles, triples, and quadruples weights for the \ce{LiH} molecule obtained with the cc-pVTZ basis set. The equilibrium
    bond distance is $R_e=3.037\,\text{a}_0$. Energy differences are reported in $\text{mE}_\text{h}$ and the FCI energies are
    $-8.03664666\,\text{E}_\text{h}$ at $R_e$, $-7.96676083\,\text{E}_\text{h}$ at $2R_e$, and $-7.94676936\,\text{E}_\text{h}$ at $3R_e$.  
    \label{tab:mp2_cc2_ccsd_cc3_ccsd_t_weights_LiH}}
\begin{tabular}{S[table-format=1.1] S[table-format=2.5] S[table-format=2.5] S[table-format=2.5] S[table-format=2.5] S[table-format=2.5] S[table-format=2.5] S[table-format=2.5]}
\hline 
\hline 
 {$R/R_e$}	& {MP2}    & {CC2}    & {CCSD}   & {CCSD(T)} & {CC3}    & {CCSDT}  & {FCI}    \\
\hline
    {$W_0$} &          &          &          &           &          &          &          \\
 1  	    &  0.98383 &  0.98359 &  0.96855 &  0.96840  &  0.96841 &  0.96837 &  0.96842 \\
 2  	    &  0.96921 &  0.94258 &  0.82731 &  0.82316  &  0.82498 &  0.82440 &  0.82456 \\
 3  	    &  0.91068 &  0.62084 &  0.39707 &  0.33539  &  0.39154 &  0.39024 &  0.39100 \\
    {$W_1$} &          &          &          &           &          &          &          \\
 1  	    &  0.00000 &  0.00015 &  0.00040 &  0.00041  &  0.00041 &  0.00041 &  0.00041 \\
 2  	    &  0.00000 &  0.01732 &  0.05577 &  0.05806  &  0.05692 &  0.05720 &  0.05719 \\
 3  	    &  0.00000 &  0.19444 &  0.29819 &  0.34268  &  0.30134 &  0.30207 &  0.30174 \\
    {$W_2$} &          &          &          &           &          &          &          \\
 1  	    &  0.01617 &  0.01626 &  0.03105 &  0.04317  &  0.03117 &  0.03120 &  0.03110 \\
 2  	    &  0.03079 &  0.04010 &  0.11691 &  0.11874  &  0.11801 &  0.11829 &  0.11794 \\
 3  	    &  0.08932 &  0.18472 &  0.30474 &  0.32155  &  0.30690 &  0.30740 &  0.30623 \\
    {$W_3$} &          &          &          &           &          &          &          \\
 1  	    &  0.00000 &  0.00000 & 0.00000  &  0.00001  &  0.00001 &  0.00002 &  0.00002 \\
 2  	    &  0.00000 &  0.00000 & 0.00000  &  0.00004  &  0.00009 &  0.00011 &  0.00013 \\
 3  	    &  0.00000 &  0.00000 & 0.00000  &  0.00038  &  0.00023 &  0.00029 &  0.00057 \\
    {$W_4$} &          &          &          &           &          &          &          \\
 1  		&  0.00000 &  0.00000 & 0.00000  &  0.00000  &  0.00000 &  0.00000 &  0.00005 \\
 2  		&  0.00000 &  0.00000 & 0.00000  &  0.00000  &  0.00000 &  0.00000 &  0.00017 \\
 3  		&  0.00000 &  0.00000 & 0.00000  &  0.00000  &  0.00000 &  0.00000 &  0.00045 \\
{$\Delta E$}&          &          &          &           &          &          &          \\
 1          & 10.719   & 10.590   & 0.082    & 0.014     & 0.015    & 0.0003   &          \\
 2          & 20.333   & 18.484   & 0.201    & 0.006     & 0.043    & 0.004    &          \\
 3          & 46.820   & 25.834   & 0.644    & -1.196    & 0.123    & 0.008    &          \\
\hline
\hline
\end{tabular}
\end{table*}

\begin{table*}[htb]
\centering
\caption{Reference, singles, doubles, triples, and quadruples weights for the \ce{HF} molecule obtained with the cc-pVDZ basis set. The equilibrium
    bond distance is $R_e=1.737\,\text{a}_0$. Energy differences are reported in $\text{mE}_\text{h}$ and the FCI energies are
    $-100.23059429\,\text{E}_\text{h}$ at $R_e$, $-100.06493232\,\text{E}_\text{h}$ at $2R_e$, and $-100.03732519\,\text{E}_\text{h}$ at $2.5R_e$.
    \label{tab:mp2_cc2_cc3_ccsd_t_weights_HF}}
\begin{tabular}{S[table-format=1.1] S[table-format=2.5] S[table-format=2.5] S[table-format=2.5] S[table-format=2.5] S[table-format=2.5] S[table-format=2.5] S[table-format=2.5]}
\hline 
\hline 
    {$R/R_e$}& {MP2}     & {CC2}     & {CCSD}     & {CCSD(T)}     & {CC3} & {CCSDT} & {FCI} \\
\hline
    {$W_0$} &          &          &          &           &          &          &          \\
1.0	    &  0.96013      &  0.95912 	  &  0.95755      &  0.95620      &  0.95606 &  0.95607 &	0.95654 \\
2.0	    &  0.90859      &  0.88232 	  &  0.82333      &  0.77994      &  0.79469 &  0.79313 &	0.79455 \\
2.5	    &  0.84137      &  0.77753 	  &  0.66624      &  0.51708      &  0.60795 &  0.61105 &	0.61869 \\
    {$W_1$} &          &          &          &           &          &          &          \\
1.0	    &  0.00000      &  0.00048 	  &  0.00038 	  &  0.00036 	  &  0.00040 &  0.00041 &	0.00040 \\
2.0	    &  0.00000      &  0.01260 	  &  0.02499      &  0.03443      &  0.02981 &  0.02992 &	0.03021 \\
2.5	    &  0.00000      &  0.02923 	  &  0.06599      &  0.10909      &  0.08013 &  0.07704 &	0.07593 \\
    {$W_2$} &          &          &          &           &          &          &          \\
1.0	    &  0.03987      &  0.04040 	  &  0.04207 	  &  0.04317 	  &  0.04327 &  0.04324 &	0.04196 \\
2.0	    &  0.09141      &  0.10507 	  &  0.15168      &  0.18375      &  0.17358 &  0.17471 &	0.16928 \\
2.5	    &  0.15863      &  0.19324 	  &  0.26776      &  0.36897      &  0.30813 &  0.30761 &	0.29451 \\
    {$W_3$} &          &          &          &           &          &          &          \\
1.0	    &  0.00000  	&  0.00000    &  0.00000	  &  0.00027 	  &  0.00027 &  0.00029 &	0.00027\\
2.0	    &  0.00000  	&  0.00000    &  0.00000  	  &  0.00187      &  0.00192 &  0.00224 &	0.00235 \\
2.5	    &  0.00000  	&  0.00000    &  0.00000  	  &  0.00486      &  0.00379 &  0.00430 &	0.00492 \\
    {$W_4$} &          &          &          &           &          &          &          \\
1.0		&  0.00000  	&  0.00000    &  0.00000      &  0.00000  	  &  0.00000 &  0.00000 &  0.00058 \\
2.0		&  0.00000  	&  0.00000    &  0.00000      &  0.00000  	  &  0.00000 &  0.00000 &  0.00285 \\
2.5		&  0.00000  	&  0.00000    &  0.00000      &  0.00000  	  &  0.00000 &  0.00000 &  0.00480 \\
 {$\Delta E$} &          &          &          &           &          &          &          \\
1.0 & 7.391 & 6.644 & 2.432 & 0.491 & 0.402 & 0.407 & \\
2.0 & 27.398 & 19.085 & 10.329 & 0.321 & 1.611 & 1.214 & \\
2.5 & 46.719 & 28.342 & 17.444 & -6.566 & 1.957 & 1.349 & \\
\hline  
\hline
\end{tabular}
\end{table*}

Table \ref{tab:mp2_cc2_ccsd_fci_H2} shows that the CCSD and FCI weights agree for the \ce{H2} molecule, also at stretched bond lengths, as they should for a two-electron system.
At $6R_e$, the doubles weight is dominated by the $\ket{\sigma_u^2}$ configuration and is roughly equal to the $\ket{\sigma_g^2}$ reference weight, as expected.
The MP2 and CC2 weights are excellent approximations at $R_e$ but quickly deteriorate as the bond length is increased. This is caused by the diminishing
gap between the occupied $\sigma_g$ orbital and the virtual $\sigma_u$ orbital at stretched bond lengths, causing overstimation of the dominant doubles
amplitudes by the second-order perturbation treatment. This is also reflected in the energy errors, which initially increase with $R$ and subsequently
decrease such that the energy eventually falls below the FCI one. This is an archetypical failure of perturbation theory.

For the \ce{LiH} molecule the CCSD and FCI energies and weights are in very good agreement, see Table \ref{tab:mp2_cc2_ccsd_cc3_ccsd_t_weights_LiH}.
The reference weight, corresponding to the configuration $\ket{1\sigma^22\sigma^2}$, is $0.969$
at the equilibrium distance $R_e$, decaying to $0.823$ and $0.397$ at $2R_e$ and $3R_e$, respectively.
At the stretched geometries, there are significant contributions from both singles and doubles, while the triples and quadruples weights
remain small ($<0.001$) and essentially negligible.
The singles weight mainly comes from the configuration
$\ket{1\sigma^2 2\sigma 3\sigma}$ with a weight of $0.046$ at $2R_e$ and $0.280$ at $3R_e$ in the FCI wave function.
The corresponding CCSD singles weight is $0.045$ at $2R_e$ and $0.277$ at $3R_e$.
In the FCI wave function, the dominating double-excited configurations are
$\ket{1\sigma^2 3\sigma^2}$ and
$\ket{1\sigma^2 3\sigma 4\sigma}$
with weights of $0.017$ and $0.027$ at $2R_e$, and $0.162$ and $0.086$ at $3R_e$ in the FCI wavefunction. The corresponding CCSD doubles weights are $0.016$ and $0.027$
at $2R_e$, and $0.161$ and $0.085$ at $3R_e$.

The MP2 and CC2 approximations overestimate the reference weight with a concomitant underestimation of the doubles weight. This is also reflected in the energy errors
which are two orders of magnitude greater than the CCSD ones. The CC3 method performs somewhat better than the CCSD(T) approximation, with results closer to the CCSDT
and FCI ones. In particular, the CCSD(T) energy falls below the FCI one at $3R_e$ while the CC3 energy remains above. The CCSDT energies agree with the FCI ones to within
a few $\mu\text{E}_\text{h}$ at all distances. While triples weights are insignificant, the triples amplitudes clearly influence the reference, singles, and doubles weights,
improving the already good agreement with FCI weights at the CCSD level.

Somewhat larger deviations are observed for the \ce{HF} molecule in Table \ref{tab:mp2_cc2_cc3_ccsd_t_weights_HF},
especially at stretched geometries. The RHF refence configuration
$\ket{1\sigma^2 2\sigma^2 1\pi^4 3\sigma^2}$ dominates with a weight slightly below $96\%$ at the equilibrium distance. At the stretched geometries, the FCI and CCSD methods agree
that two excited configurations---the
single-excited $\ket{1\sigma^2 2\sigma^2 1\pi^4 3\sigma 4\sigma}$ and the double-excited $\ket{1\sigma^2 2\sigma^2 1\pi^4 4\sigma^2}$---contribute
significantly. Their weights at $2R_e$ are $0.026$ and $0.131$ in the FCI wave function, while the CCSD method predicts $0.021$ and $0.110$.
At $2.5R_e$, the FCI and CCSD weights are $0.068$, $0.267$ and $0.059$, $0.233$, respectively.
Also for the \ce{HF} molecule, the quality of the second-order approximations decrease as the bond is stretched. The CC3 method performs better than
the CCSD(T) approximation, especially at stretched geometries. Although the triples and quadruples weights are small ($<0.005$), the inclusion of triples
in the cluster operators improves the reference, singles, and doubles weights. Overall, therefore, these preliminary investigations indicate a hierarchy
of weight approximations following the order $\text{MP2} < \text{CC2} < \text{CCSD} < \text{CCSD(T)} < \text{CC3} < \text{CCSDT}$. The apparent superiority
of the CC3 method over the CCSD(T) approximation is not too surprising, of course, since the latter is aimed at a perturbative correction of the
energy while the former is a similar correction of the wave function.

It is well known that the CCSD method works well for the systems considered above, at least in terms of the energy.
Our investigation shows that the CCSD weights also are good approximations to the FCI weights for these systems.
To challenge the CC weight concept, we now turn our attention to the \ce{N2} molecule, which is single-reference dominated at the
equilibrium bond distance and rapidly develops increasing multi-reference character as the bond is stretched.
This should be clearly reflected
in the CCSD weights deviating substantially from FCI results as the bond is stretched. Indeed, this is what we observe from the
data presented in Table \ref{tab:N2_ccsd_qccsd_fci} where we have also included results obtained with the CC2, QCCSD, CCSD(T), CC3, and CCSDT models for comparison.
\begin{table*}[htb]
\centering
\caption{Reference, singles, doubles, triples, and quadruples weights for the \ce{N2} molecule obtained with the 6-31G basis set.
    The equilibrium bond distance is $R_e=2.102\,\text{a}_0$. Energy differences are reported in $\text{mE}_\text{h}$ and the FCI energies are
    $-109.10719404\,\text{E}_\text{h}$ at $R_e$, $-109.00405250\,\text{E}_\text{h}$ at $1.3R_e$, and $-108.89416902\,\text{E}_\text{h}$ at $1.6R_e$.\label{tab:N2_ccsd_qccsd_fci}}
\begin{tabular}{S[table-format=1.1] S[table-format=2.5] S[table-format=2.5] S[table-format=2.5] S[table-format=2.5] S[table-format=2.5] S[table-format=2.5] S[table-format=2.5]}
\hline 
\hline 
{$R/R_e$}& {CC2}& {CCSD}   & {QCCSD}  & {CCSD(T)} & {CC3}    & {CCSDT}  & {FCI}    \\
\hline
{$W_0$}&        &          &          &           &          &          &          \\
 1.0 &  0.88888 &  0.89993 &  0.90053 &  0.89100  &  0.89036 &  0.89107 &  0.89218 \\
 1.3 &  0.70313 &  0.79704 &  0.80077 &  0.76451  &  0.76658 &  0.77094 &  0.78069 \\
 1.6 &  0.25675 &  0.33220 &  0.57363 &  0.09715  &  0.28670 &  0.21930 &  0.53554 \\
{$W_1$}&        &          &          &           &          &          &          \\
 1.0 &  0.00386 &  0.00217 &  0.00173 &  0.00179  &  0.00202 &  0.00205 &  0.00205 \\
 1.3 &  0.01866 &  0.00574 &  0.00413 &  0.00435  &  0.00562 &  0.00578 &  0.00562 \\
 1.6 &  0.05806 &  0.01245 &  0.00697 &  0.01054  &  0.00862 &  0.00872 &  0.00856 \\
{$W_2$}&        &          &          &           &          &          &          \\
 1.0 &  0.10727 &  0.09790 &  0.09358 &  0.10516  &  0.10544 &  0.10461 &  0.09852 \\
 1.3 &  0.27820 &  0.19722 &  0.17828 &  0.22478  &  0.22090 &  0.21782 &  0.18699 \\
 1.6 &  0.68519 &  0.65536 &  0.33049 &  0.87494  &  0.68799 &  0.75514 &  0.36339 \\
{$W_3$}&        &          &          &           &          &          &          \\
 1.0 &  0.00000 &  0.00000 &  0.00013 &  0.00205  &  0.00219 &  0.00227 &  0.00202 \\
 1.3 &  0.00000 &  0.00000 &  0.00068 &  0.00637  &  0.00690 &  0.00547 &  0.00447 \\
 1.6 &  0.00000 &  0.00000 &  0.00363 &  0.01737  &  0.01669 &  0.01684 &  0.00922 \\
{$W_4$}&        &          &          &           &          &          &          \\
 1.0 &  0.00000 &  0.00000 &  0.00403 &  0.00000  &  0.00000 &  0.00000 &  0.00495 \\
 1.3 &  0.00000 &  0.00000 &  0.01614 &  0.00000  &  0.00000 &  0.00000 &  0.02025 \\
 1.6 &  0.00000 &  0.00000 &  0.08529 &  0.00000  &  0.00000 &  0.00000 &  0.07392 \\
{$\Delta E$}&   &          &          &           &          &          &          \\
 1.0 & -7.206   & 9.860    & 7.858    & 1.925     & 1.459    & 2.021    &          \\
 1.3 & -67.969  & 24.939   & 18.441   & 4.891     & 2.926    & 7.283    &          \\
 1.6 & -199.877 & 36.411   & 29.944   & -10.146   & -4.053   & 2.765    &          \\
\hline
\hline
\end{tabular}
\end{table*}

We first note that the FCI weights up to quadruples sum to $0.99972$ at $R_e$, $0.99802$ at $1.3R_e$, and $0.99063$ at $1.6R_e$ and, thus,
excited determinants beyond quadruples contribute less than $1\%$ at all three distances.
The RHF reference determinant is $\ket{1\pi^45\sigma^2}$ where, for convenience, we have included only the highest-lying occupied orbitals in the notation.
It has a weight of $0.892$ in the FCI wave function at the equilibrium distance, dropping rapidly to $0.781$ and $0.536$ at $1.3R_e$ and $1.6R_e$, respectively.
The dominant double-excited configuration in the FCI wave function is $\ket{1\pi^25\sigma^22\pi^2}$
with a weight of $0.025$ ($25\%$ of $W_2$) at the equilibrium distance, increasing to $0.042$ ($22\%$) at 
$1.3R_e$ and $0.173$ ($48\%$) at $1.6R_e$.
Also, at $1.6R_e$ the quadruple-excited configuration $\ket{5\sigma^22\pi^4}$ becomes non-negligible with a weight of $0.027$ ($36\%$ of $W_4$) in the FCI wave function.
At $R_e$ and $1.3R_e$ the CCSD and QCCSD weights are in reasonably good agreement with the FCI weights. At $1.6R_e$, however, the CCSD method severely underestimates
the FCI reference weight and overestimates the doubles weight, indicating a failure of the CCSD method despite an energy error on the same order of magnitude as
at the shorter bond distances.
On the other hand, the QCCSD model only modifies the bra state compared with the CCSD model
and provides a much-improved approximation of the FCI weights with roughly the same energy errors.
This can be attributed to the fact that disconnected triples and quadruples are included in $\bra{\tilde{\Psi}^{\text{QCCSD}}}$.
These contribute not only to the triples and quadruples weights but also to the reference, singles, and doubles weights. In addition, the singles and doubles
amplitudes are indirectly affected by the quadratic term of the bra through the amplitude equations.

The CC2 weights are quite similar to the CCSD ones, deviating significantly from the FCI ones as the \ce{N2} bond is elongated, but with much greater
energy errors, all \emph{below} the FCI energy.
Including triples in the description improves the energy but does not improve the agreement for the weights.
In fact, the reference and doubles weights are even further from the
FCI ones than the CCSD weights, especially for the CCSD(T) model, although the CCSD(T), CC3, and CCSDT energies agree to within $\sim 10\,\text{mE}_\text{h}$
at all three distances.

\subsection{Effect of orbital choice}

Unlike FCI theory, truncated CC models rely on a reference determinant that is close enough to the FCI wave function.
It is well known, however, that the CCSD model can compensate for a poor choice of reference determinant through the
approximate orbital relaxation provided by the $\exp(\hat{T}_1)$ operator. This makes the CCSD model (and higher-order truncated CC models)
near-invariant to the choice of reference determinant, i.e., to the choice of orbital basis.
One typically chooses the HF determinant which may be a poor choice at, e.g., stretched bond lengths.
In cases where the HF solution shows pathological behavior, one may try other choices such as the KS determinant and rely on $\exp(\hat{T}_1)$ to approximately rotate the reference determinant into a closer-to-optimal one.
The NOCCD model includes a complete biorthonormal orbital rotation and, in essence, thus defines a new reference determinant to which the CCD approach is applied. It should be stressed, of course, that the new reference determinant of the NOCCD model is determined in concert with the correlation.
These effects can be illustrated by the CC weight concept.

\begin{figure}[h]
    \includegraphics[scale=0.75]{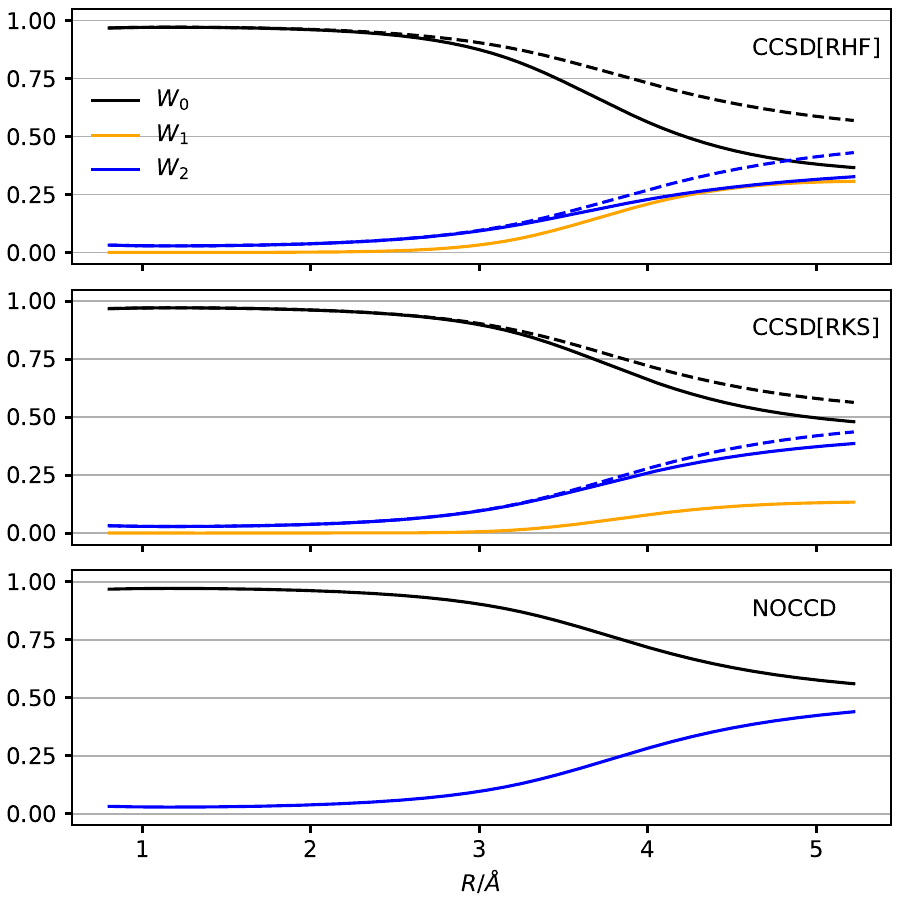}
    \caption{$W_0$, $W_1$, and $W_2$ for the \ce{LiH} molecule as functions of the bond distance. The top two panels show results
    obtained from CCSD theory using the RHF and RKS reference determinants; full lines: bare determinant basis, dashed lines: 
    $\hat{T}_1$-transformed basis. The last panel shows results obtained
    in the fully rotated determinant basis with NOCCD theory. The cc-pVTZ basis set is used in all calculations, and smooth curves are obtained by cubic spline interpolation.}
  \label{fig:weights_LiH_cc-pvtz}
\end{figure}
We first consider the \ce{LiH} molecule for which the CCSD model provides an excellent approximation of the FCI energy across the ground-state
potential-energy curve, despite a significant reduction of the reference weight at stretched bond lengths. Figure \ref{fig:weights_LiH_cc-pvtz}
shows $W_0$, $W_1$, and $W_2$ obtained with the CCSD model using either the RHF reference (denoted CCSD[RHF]) or the KS reference (denoted CCSD[RKS])
with the TPSS0 density-functional approximation. 
For these methods, the weights are computed by projection onto the bare determinants using $\hat{P}_1$ and by projection
onto the $\hat{T}_1$-transformed determinants using $\exp(\hat{T}_1)\hat{P}_1 \exp(-\hat{T}_1)$. Finally, we also show the reference and doubles weights
obtained from NOCCD theory by projection onto the rotated determinants using $\exp(\hat{\kappa})\hat{P}_1 \exp(-\hat{\kappa})$.

The potential-energy curves obtained from the CCSD[RHF], CCSD[RKS], and NOCCD models are nearly identical,
indicating the approximate orbital invariance.
At the equilibrium bond distance, $R_e = 1.596\,\text{\AA}$, the CCSD[RKS] and NOCCD energies are $1.04\,\mu\text{E}_\text{h}$ ($2.74\,\text{J/mol}$)
and $41.5\,\mu\text{E}_\text{h}$ ($109\,\text{J/mol}$) above the CCSD[RHF] energy. The maximum deviation across the potential-energy curves is
$44.2\,\mu\text{E}_\text{h}$ ($116\,\text{J/mol}$) for the CCSD[RKS] method and $210\,\mu\text{E}_\text{h}$ ($552\,\text{J/mol}$) for the NOCCD method with respect to the
CCSD[RHF] approximation.

As seen in Fig.~\ref{fig:weights_LiH_cc-pvtz}, the reference weight is close to unity for distances up to about
$2.5\,\text{\AA}$. At greater distances, the reference weight drops, falling below $50\%$ for both the CCSD[RHF]
and CCSD[RKS] methods. With the CCSD[RHF] model, the weight is transferred roughly equally to 
$W_1$ and $W_2$, indicating significant approximate orbital relaxation due to $\exp(\hat{T}_1)$.
Indeed, the $\hat{T}_1$-transformed reference weight is substantially greater than the untransformed one.
The same effect is observed with the CCSD[RKS] model, although much less pronounced.
The singles weight increases but much less than in the CCSD[RHF] case. As one might perhaps have
expected, the $\hat{T}_1$-transformed reference and doubles weights are roughly the same for the
CCSD[RHF] and CCSD[RKS] models. 
With mean absolute deviations of $0.02$ for $W_0$ and $0.01$ for $W_2$,
the NOCCD model predicts weights that closely agree with
the $\hat{T}_1$-transformed weights of the CCSD[RHF] and CCSD[RKS] theories.
Hence, the CCSD[RHF], CCSD[RKS], and NOCCD approximations 
provide the same qualitative picture of the correlated
ground state of \ce{LiH} across the potential-energy surface.

\begin{figure}[h]
    \includegraphics[scale=0.75]{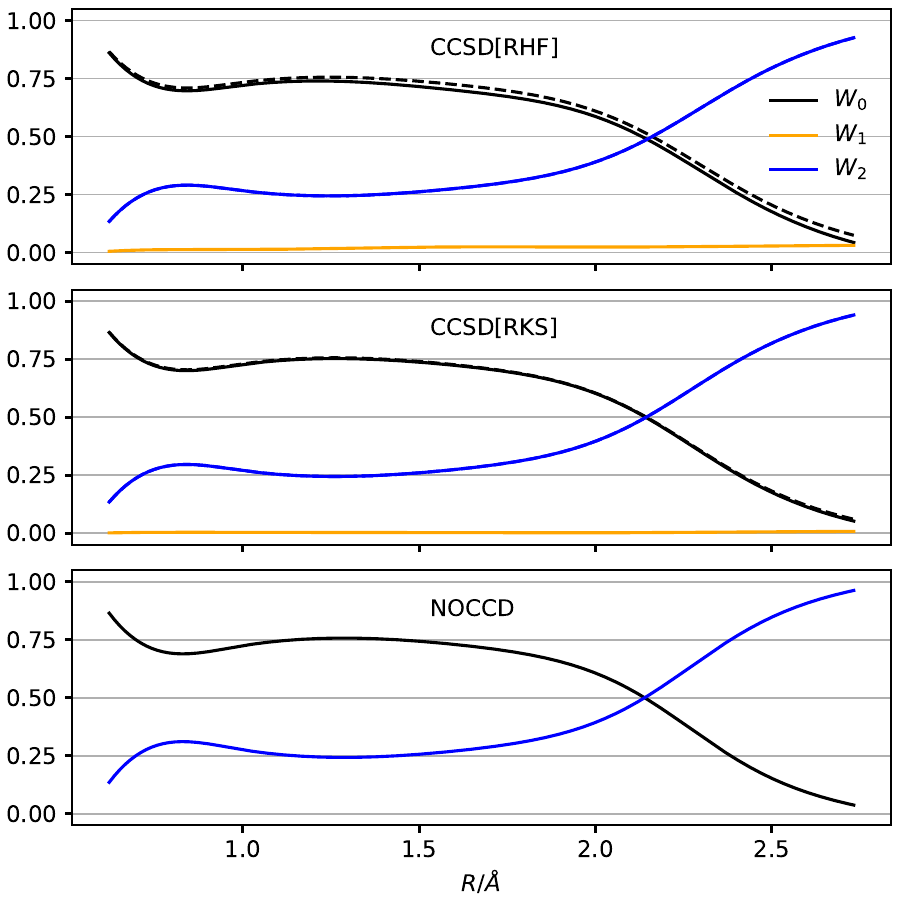}
    \caption{Same as Fig.~\ref{fig:weights_LiH_cc-pvtz}, here for the \ce{C2} molecule.}
  \label{fig:weights_C2_cc-pvtz}
\end{figure}
We next turn to the \ce{C2} molecule where the CCSD approximation fails due to multi-reference character.
The calculations are done in the same way as the \ce{LiH} ones above, and with the same basis set (cc-pVTZ).
Some deviations are seen already at the CCSD[RHF] equilibrium bond distance $R_e = 1.242\,\text{\AA}$ where the
CCSD[RKS] approach predicts an energy $2.42\,\text{mE}_\text{h}$ ($6.34\,\text{kJ/mol}$) above the CCSD[RHF] energy.
The NOCCD energy is somewhat higher, at $2.53\,\text{mE}_\text{h}$ ($6.63\,\text{kJ/mol}$) above the CCSD[RHF] energy.
At $R=2.732\,\text{\AA}$, the CCSD[RKS] energy is $0.15\,\text{mE}_\text{h}$ ($0.39\,\text{kJ/mol}$) below the CCSD[RHF] one,
while the NOCCD energy is $2.30\,\text{mE}_\text{h}$ ($6.03\,\text{kJ/mol}$) below.

The weights are plotted in Fig.~\ref{fig:weights_C2_cc-pvtz}. At $R_e$, the CCSD[RHF] approximation predicts a
reference weight of roughly $75\%$, with the remaining $25\%$ residing mainly in double-excited determinants and very little in single-excited determinants. This picture is also obtained with the RKS reference,
albeit with even less population in the single-excited determinants. Also the NOCCD method agrees.
The singles weight increases a bit in the CCSD[RHF] state as the bond length is increased, whereas it
remains negligible at all bond distances in the CCSD[RKS] state. Hence, the $\exp(\hat{T}_1)$ operator can
do only little to improve the reference. All three methods agree that the reference weight nearly vanishes
at $R=2.732\,\text{\AA}$, with the doubles weight reaching close to $100\%$. This, of course, indicates a strongly
correlated system, although one must always keep in mind the orbital-dependence of the weights.

Regardless of the reference choice, at $R=2.732\,\text{\AA}$,
the CCSD doubles weight is dominated by four distinct combinations of $\pi$--$\pi^*$ (HOMO--LUMO)
double excitations. One of them turns out to be negative, $-0.113$ with the RKS reference and
$-0.106$ with the RHF reference. Such out-of-bounds weights can be taken as an indication that the
state is poorly described with the CCSD approximation (keeping in mind the inherent orbital-dependence,
of course). Along these lines, it should be noted that the total $W_0$ and $W_2$ would become negative
and greater than $1$, respectively, if the bond distance is further increased in
Fig.~\ref{fig:weights_C2_cc-pvtz}, for all three methods. This merely illustrates
that one cannot remedy multi-reference character by choosing a single reference determinant.

Large doubles amplitudes $\tau_{ij}^{ab}$ have long been taken as an indication of strong correlation or
potential multi-reference character, although the precise and general definition of ``large'' remains unclear.
It is interesting to note that the doubles amplitudes corresponding
to the dominant $\pi$--$\pi^*$ doubles weights for \ce{C2} at $R=2.732\,\text{\AA}$ are also
by far the largest amplitudes, accounting for more than $80\%$ of the total (Frobenius)
norm of the entire amplitude array. However, the amplitude corresponding to the determinant with the
greatest weight is not the one with the greatest amplitude value. In fact, it only accounts for about
$10\%$ of the total amplitude norm, illustrating the difficulties faced when trying to define the precise
meaning of ``large'' doubles amplitudes.

\subsection{Noninteracting subsystems}

To elucidate the separability issues associated with the linear parameterization of $\bra{\tilde{\Psi}}$,
we consider the $\ce{H2}$ dimer. The two hydrogen molecules both have bond distance $R$ and are placed in a
parallel configuration with a separation denoted $D$. That is, the four protons form a rectangle with
side lengths $R$ and $D$. Choosing $D = 1000\,\text{a}_0$, the two hydrogen molecules can be considered noninteracting.

By size-consistency, the CCSD energy of the dimer will be equal to twice the monomer energy. Since \ce{H2} is
a two-electron system, the monomer energy will be equal to the exact result, the FCI energy.
Hence, the CCSD energy of the dimer will be exact for all values of $R$. Due to the linear parameterization of
$\bra{\tilde{\Psi}}$, however, the bivariational CCSD ground state
of the dimer is not exact, and separability issues are expected to arise in the weights.
Since the \ce{H2} molecules are effectively noninteracting,
the components missing in the linear $\hat{\Lambda}$ operator are disconnected triples and quadruples.
These are included in the QCCSD model, albeit in an approximate fashion. Hence, the QCCSD model
should yield both the exact energy \emph{and} a much-improved approximation of the weights.

For reference, Table~\ref{tab:H2_fci_ccpvdz} contains the FCI energies and weights obtained for
the \ce{H2} molecule with the cc-pVDZ basis set.
\begin{table*}[h]
    \centering
    \caption{Reference, singles, and doubles weights for the \ce{H2} molecule obtained from the FCI wave function
    with the cc-pVDZ basis. The equilibrium bond distance is $R_e = 1.4\,\text{a}_0$.\label{tab:H2_fci_ccpvdz}}
\begin{tabular}{l S[table-format=2.8] S[table-format=1.5] S[table-format=1.5]  S[table-format=1.5] S[table-format=2.5] S[table-format=1.5] S[table-format=2.5]}
    \hline
    \hline
    {$R/R_e$} &{$E$/$\text{E}_h$}& {$W_0$}  & {$W_1$} & {$W_2$} \\
    \hline
    1    & -1.16339873 & 0.98311 & 0.00010 & 0.01678  \\
    2    & -1.06392796 & 0.91291 & 0.00268 & 0.08441  \\
    4    & -0.99966961 & 0.56362 & 0.01551 & 0.42086  \\
    \hline
    \hline
\end{tabular}
\end{table*}
The energies and weights obtained from the CCSD (and QCCSD) method are identical and, therefore, not
reported. Using Eqs.(\ref{eq:W0_sep})-(\ref{eq:W2_sep}), we can easily predict the reference, singles,
and doubles weights that should be obtained for the noninteracting dimer.

Our results for the \ce{H2} dimer are reported in Table~\ref{tab:H2_dimer_ccsd_qccsd_fci}.
\begin{table*}[h]
\centering
    \caption{Reference, singles, doubles, triples, and quadruples weights for two noninteracting
    \ce{H2} molecules (separated by $1000\,\text{a}_0$). The monomer equilibrium bond distance is
    $R_e=1.4\,\text{a}_0$. All results are obtained with the cc-pVDZ basis set.
    \label{tab:H2_dimer_ccsd_qccsd_fci}}
\begin{tabular}{l S[table-format=1.0] S[table-format=1.5] S[table-format=1.5] S[table-format=1.5]}
\hline 
\hline 
      &{$R/R_e$}&{CCSD}       &{QCCSD}        &{FCI}\\
\hline
$W_0$ & 1	    &  0.96622 	  &  0.96651 	  &  0.96651	 \\
	  & 2       &  0.82582	  &  0.83340	  &  0.83340	 \\
	  & 4       &  0.12721 	  &  0.31773      &  0.31767 	 \\
$W_1$ & 1	    &  0.00021 	  &  0.00021 	  &  0.00021	 \\
	  & 2     	&  0.00536 	  &  0.00489 	  &  0.00489	 \\
	  & 4     	&  0.03109 	  &  0.01737 	  &  0.01749	 \\
$W_2$ & 1	    &  0.03357 	  &  0.03300 	  &  0.03300	 \\
	  & 2    	&  0.16883 	  &  0.15413 	  &  0.15413	 \\
	  & 4     	&  0.84170 	  &  0.47465 	  &  0.47466	 \\
$W_3$ & 1	    &  0.00000    &  0.00000 	  &  0.00000	 \\
	  & 2     	&  0.00000    &  0.00045 	  &  0.00045	 \\
	  & 4     	&  0.00000    &  0.01316 	  &  0.01306	 \\
$W_4$ & 1	    &  0.00000    &  0.00028 	  &  0.00028	 \\
	  & 2     	&  0.00000    &  0.00713 	  &  0.00713	 \\
	  & 4     	&  0.00000    &  0.17708 	  &  0.17713	 \\
\hline 
\hline
\end{tabular}
\end{table*}
The energies obtained from the CCSD, QCCSD, and FCI methods are identical and equal to twice the monomer
energies reported in Table~\ref{tab:H2_fci_ccpvdz}, as required by size-consistency.
It is easily verified that the FCI reference, singles, and doubles weights exactly satisfy
Eqs.~\eqref{eq:W0_sep}-\eqref{eq:W2_sep}.

For the CCSD method, however, we observe deviations due the lack of multiplicative separability of
$\bra{\tilde{\Psi}}$. While the deviations are almost negligible at $R=R_e$, they
increase rapidly with $R$, and at $R=4R_e$, the reference, singles, and doubles
weights are off by roughly a factor of $2$.

The QCCSD method yields a significant improvement, almost exactly reproducing the FCI results for $W_0$, $W_1$,
and $W_2$ at all $R$. The greatest deviation is on the order of $10^{-4}$ for the reference and singles weights
at $R=4R_e$. With the QCCSD method we can also compare the triples and quadruples weights with the FCI results.
Also for these, we observe an excellent agreement.

As mentioned above, the CCSD method provides an excellent approximation to the FCI energy for the \ce{LiH} molecule.
If we consider the \ce{LiH} dimer in a noninteracting rectangular configuration analogous to the one used
for the \ce{H2} dimer above, the CCSD dimer energy remains accurate thanks to size-consistency. The
\ce{LiH} molecule, however, is a four-electron system and the CCSD \emph{Ansatz} is not formally exact. The
two core electrons only contribute very little to the electron correlation energy as the bond distance is increased and,
consequently, the \ce{LiH} molecule can be seen as \emph{almost} a two-electron system in this context.

\begin{figure}[h]
    \includegraphics[scale=0.75]{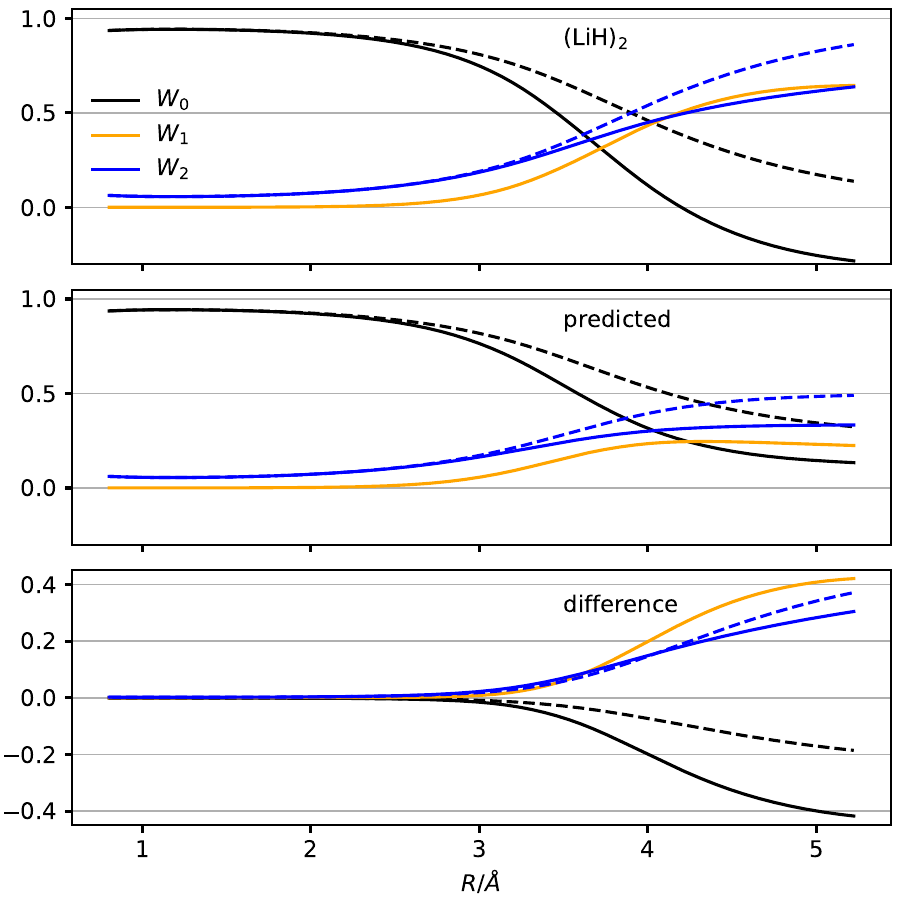}
    \caption{Weights computed for the noninteracting \ce{LiH} dimer,
    presented as in Fig.~\ref{fig:weights_LiH_cc-pvtz}.
    The top panel shows results obtained with the CCSD method using the RHF reference and the cc-pVTZ basis set.
    The middle panel shows the results predicted by Eqs.~\eqref{eq:W0_sep}-\eqref{eq:W2_sep} using the
    monomer data in the top panel of Fig.~\ref{fig:weights_LiH_cc-pvtz}.
    The last panel shows the difference between the two.}
    \label{fig:weights_LiH_dimer_cc-pvtz}
\end{figure}

We present CCSD weights for the \ce{LiH} dimer as functions of the \ce{Li}--\ce{H} distance $R$
in Fig.~\ref{fig:weights_LiH_dimer_cc-pvtz}.
The behavior of the weights as functions of $R$ is qualitatively similar to the monomer case presented
in Fig.~\ref{fig:weights_LiH_cc-pvtz}, but we immediately notice that the reference weight becomes negative at
distances beyond roughly $4\,\text{\AA}$. The $\hat{T}_1$-transformed weights remain within bounds, however, at least
at the distances considered.

Using Eqs.~\eqref{eq:W0_sep}-\eqref{eq:W2_sep} to predict the dimer weights clearly does not produce
negative weights at large distances. Rather, the predicted reference, singles, and doubles weights 
appear to converge to values well within bounds at large $R$. 
The difference between the computed and predicted dimer weights are negligible or small for distances up
to about twice the \ce{LiH} equilibrium distance $R_e = 1.5958\,\text{\AA}$, however. 

\begin{figure}[h]
    \includegraphics[scale=0.75]{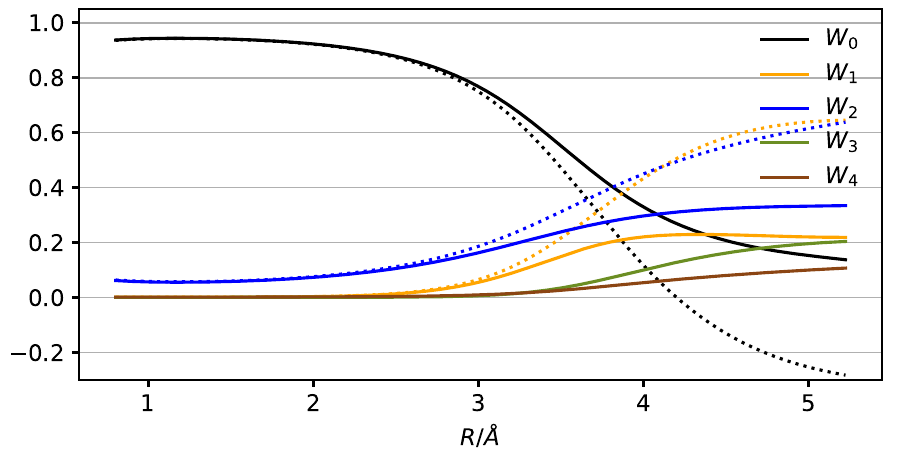}
    \caption{QCCSD reference, singles, doubles, triples, and quadruples weights computed for the
    noninteracting \ce{LiH} dimer with the cc-pVTZ basis set. The dotted lines show the CCSD weights
    for comparison.}
    \label{fig:weights_LiH_dimer_cc-pvtz_qccsd}
\end{figure}

As can be seen in Fig.~\ref{fig:weights_LiH_dimer_cc-pvtz_qccsd}, the weights obtained with the
QCCSD method remain within bounds, in marked contrast to the conventional CCSD method.
As above, this is due to the disconnected triples and quadruples de-excitations, which cause
significant triples and quadruples weights at large \ce{Li}--\ce{H} distances
with concomitant changes in the reference, singles, and
doubles weights. The effective two-electron nature of the \ce{LiH} monomer reveals itself
through QCCSD reference, singles, and doubles weights that are almost identical to those
predicted from the CCSD monomer data in Fig.~\ref{fig:weights_LiH_cc-pvtz}. The maximum relative
deviation between the QCCSD and predicted CCSD weights are $6\%$ for the reference, $7\%$ for the singles,
and $2\%$ for the doubles.

\section{Summary and concluding remarks}

We have demonstrated that weights can be defined within CC theory as bivariational expectation values
of projection operators. This allows for a wave-function analysis analogous to configuration-interaction-based 
models for all approximate CC models, including those that are based on perturbation theory (e.g., the CCSD(T) method) and thus do
not provide an explicitly parameterized right (ket) or left (bra) wave function.
We note, however, that weights cannot be used as strict diagnostics for multi-reference character, as they are neither size-consistent
nor invariant to the choice of orbital basis. The latter applies, of course, to both CC theory and configuration-interaction-based theories, including
FCI theory. The orbital-dependence can be turned into an advantage, since the
weights nicely capture the effect of the choice of orbital basis. In particular, the well known orbital-relaxation effect of
the $\exp(\hat{T}_1)$ operator is easily seen to correct short-comings of the chosen reference determinant
in such a way that nearly the same energy is obtained with any reasonable reference.

The main disadvantage of the CC weights concept is the lack of proper separability for noninteracting
subsystems, a concept closely related to size-consistency. The culprit is the linear parameterization
of the left (bra) state, $\bra{\tilde{\Psi}}$, which breaks the multiplicative separability observed
for the FCI wave function. Only in the untruncated (full CC) limit is separability guaranteed.
Most likely, the only way to correct this behavior is to use Arponen's extended CC theory. This is
corroborated by results obtained with quadratic CC theory where we observe a much-improved behavior.

One might perhaps argue that the lack of proper separability tells one that expectation values of
operators that are not additively separable---such as the projection operators defining the CC
weights---should not be computed in the usual CC manner, i.e., using the bivariational form, Eq.~\eqref{eq:expect}, with
the linear $\hat{\Lambda}$ operator.
However, the linear operator naturally appears for both ground and excited states in the widely used
EOM-CC theory,~\cite{Geertsen1989,Comeau1993,Stanton1993}
despite the associated separability issues.~\cite{Koch1994,Nanda2018}
In addition, it should be recalled that the CC one-electron density matrix consists of elements defined
as expectation values,
\begin{equation}
    D_{pq} = \braket{\tilde{\Psi} \vert \hat{a}_p^\dagger \hat{a}_q \vert \Psi},
\end{equation}
of products of creation and annihilation operators, which are neither additively nor
multiplicatively separable. The eigenvalues of this matrix (often after symmetrization as dictated by Eq.~\eqref{eq:expect})
are interpreted as natural occupation numbers and used to, e.g.,
define indices of multi-determinant and multi-reference character.~\cite{Bartlett2020}
As an example, the natural occupation numbers for the \ce{LiH} dimer discussed above should be exactly
identical to those obtained for the monomer (repeated twice, of course). The norm of the vector
measuring the difference between computed and expected (from monomer calculations) natural occupation
numbers for the $\ce{LiH}$ dimer increases by two orders of magnitude from $R=R_e$ to $R=3.25R_e$.
(The deviations are small enough to be negligible in this case, though: $2.3 \times 10^{-9}$ at $R=R_e$ and
$1.6 \times 10^{-7}$ at $R=3.25 R_e$.)

We conclude that the weight concept appears as a useful tool for analyzing CC states, although one
needs to be aware of the separability issues and orbital-dependence (which is always an issue, also
for configuration-interaction methods). In practice, our tests indicate that
a CC calculation may be assumed to be reliable if the reference weight is close enough to unity,
especially if the reference weight is computed in the $\hat{T}_1$-transformed basis. Reference weights
further from unity may indicate multi-reference character and potential failure of the single-reference CC method. 
However, it may also be a consequence of fundamental separability issues that do not necessarily imply poor energies.
Further testing, particularly for larger systems, is clearly needed to establish the usability of the reference weight
in this regard.

The weight concept can be straightforwardly extended to EOM-CC theory,~\cite{Geertsen1989,Stanton1993,Comeau1993} thus providing a simple and systematic characterization
of excited states in terms of electron configurations. Currently, this is usually done by judging the relative magnitudes of the components
of the EOM-CC eigenvectors.
For the same reason, the weight concept can be used to interpret
CC simulations---using either TDCC or time-dependent EOM-CC theory\cite{ofstad_time-dependent_2023}---of many-electron dynamics in terms of elementary orbital transitions,
which is the language most commonly used in experimental chemistry. This includes assignment of absorption lines obtained from the Fourier transform
of the induced dipole moment.

Finally, we note that CC weights may be useful for analyzing electron-correlation effects in ground and excited states using quantities
such as the Shannon entropy from quantum information theory.~\cite{aliverti-piuri_what_2024}

\begin{acknowledgement}
This work was supported by the Research Council of Norway through its Centres of Excellence scheme, Grant No.~262695.
The calculations were performed on resources provided by Sigma2---the National Infrastructure for High Performance Computing and
Data Storage in Norway, Grant No.~NN4654K.
\end{acknowledgement}

\appendix
\section{Appendix: Algebraic CCSDT expressions in spin-orbital basis}
\renewcommand{\theequation}{A\arabic{equation}}
\setcounter{equation}{0}
\label{appendix}

In the CCSDT approximation,
\begin{align}
    &\hat{T} = \hat{T}_1+\hat{T}_2 + \hat{T}_3, \qquad \hat{\Lambda} = \hat{\Lambda}_1 + \hat{\Lambda}_2 + \hat{\Lambda}_3,
\end{align}
where, with indices $a,b,c$ denoting virtual spin orbitals and $i,j,k$ occupied ones,
\begin{align}
    &\hat{T}_1 = \sum_{ia} \tau^a_i \hat{a}_a^\dagger \hat{a}_i, &\qquad &\hat{T}_2 = \frac{1}{4} \sum_{ijab} \tau^{ab}_{ij} \hat{a}_a^\dagger \hat{a}_i \hat{a}_b^\dagger \hat{a}_j, &\qquad &\hat{T}_3 = \frac{1}{36} \sum_{ijkabc} \tau^{abc}_{ijk} \hat{a}_a^\dagger \hat{a}_i \hat{a}_b^\dagger \hat{a}_j \hat{a}_c^\dagger \hat{a}_k, \\
    &\hat{\Lambda}_1 = \sum_{ia} \lambda^i_a \hat{a}_i^\dagger \hat{a}_a, &\qquad &\hat{\Lambda}_2 = \frac{1}{4} \sum_{ijab} \lambda^{ij}_{ab} \hat{a}_j^\dagger \hat{a}_b \hat{a}_i^\dagger \hat{a}_a, &\qquad &\hat{\Lambda}_3 = \frac{1}{36} \sum_{ijkabc} \lambda^{ijk}_{abc} \hat{a}_k^\dagger \hat{a}_c \hat{a}_j^\dagger \hat{a}_b \hat{a}_i^\dagger \hat{a}_a.
\end{align}
Then, computing $c_\mu = \braket{\Phi_\mu|\Psi}$ and $\tilde{c}_\mu = \braket{\tilde{\Psi}|\Phi_\mu}$ for $\mu\in\{0,1,2,3\}$, we obtain
\begin{align}
    c_0 &= 1, \\
    c_i^a &= \tau^a_i, \\
    c_{ij}^{ab} &= \tau^{ab}_{ij} + \tau^{a}_{i} \tau^{b}_{j} - \tau^{b}_{i} \tau^{a}_{j}, \\
    c^{abc}_{ijk} &= \tau^{abc}_{ijk}
    + \tau^a_i \tau^{bc}_{jk} 
    + \tau^a_k \tau^{bc}_{ij}
    - \tau^a_j \tau^{bc}_{ik}
    + \tau^b_j \tau^{ac}_{ik} 
    - \tau^b_i \tau^{ac}_{jk} 
    - \tau^b_k \tau^{ac}_{ij} \nonumber \\
    &+ \tau^c_i \tau^{ab}_{jk} 
    + \tau^c_k \tau^{ab}_{ij} 
    - \tau^c_j \tau^{ab}_{ik}
    + \tau^a_i \tau^b_j \tau^c_k
    - \tau^a_i \tau^b_k \tau^c_j
    + \tau^a_j \tau^b_k \tau^c_i
    - \tau^a_j \tau^b_i \tau^c_k
    + \tau^a_k \tau^b_i \tau^c_j 
    - \tau^a_k \tau^b_j \tau^c_i,   
\end{align}
and
\begin{align}
    \tilde{c}_0 &= 1- \sum_{ia} \lambda^i_a  \tau^a_i     
    -  \frac{1}{4} \sum_{ijab} \lambda^{ij}_{ab}  \tau^{ab}_{ij}  
    + \frac{1}{2}  \sum_{ijab} \lambda^{ij}_{ab} \tau^a_i  \tau^b_j   \nonumber \\  
    &-  \frac{1}{36} \sum_{ijkabc} \lambda^{ijk}_{abc} \tau^{abc}_{ijk}      
    + \frac{1}{4} \sum_{ijkabc} \lambda^{ijk}_{abc}   \tau^{ab}_{ij} \tau^c_k    
    -  \frac{1}{6} \sum_{ijkabc} \lambda^{ijk}_{abc} \tau^a_i  \tau^b_j  \tau^c_k,  \\
	\tilde{c}^i_a &= \lambda^i_a 
    - \sum_{jb} \lambda^{ij}_{ab} \tau^b_j   
    - \frac{1}{4}\sum_{jkbc} \lambda^{ijk}_{abc} \tau^{bc}_{jk}    
    + \frac{1}{2}\sum_{jkbc} \lambda^{ijk}_{abc} \tau^b_j \tau^c_k,  \\
    \tilde{c}^{ij}_{ab} &= \lambda^{ij}_{ab} 
    - \sum_{kc}\lambda^{ijk}_{abc} \tau^c_k,  \\
    \tilde{c}^{ijk}_{abc} &= \lambda^{ijk}_{abc}.
\end{align}
The CCSD weights are easily obtained from these expressions by putting the triples amplitudes equal to zero.

\providecommand{\latin}[1]{#1}
\makeatletter
\providecommand{\doi}
  {\begingroup\let\do\@makeother\dospecials
  \catcode`\{=1 \catcode`\}=2 \doi@aux}
\providecommand{\doi@aux}[1]{\endgroup\texttt{#1}}
\makeatother
\providecommand*\mcitethebibliography{\thebibliography}
\csname @ifundefined\endcsname{endmcitethebibliography}
  {\let\endmcitethebibliography\endthebibliography}{}

\end{document}